\documentclass[twocolumn,prx,aps,superscriptaddress,showpacs]{revtex4-2}%,nofootinbib]{revtex4-2}
\usepackage{CJK}
\usepackage{lineno}
\usepackage{graphicx}
\usepackage{amsmath}
\usepackage{amssymb}
\usepackage{braket}
\usepackage[english]{babel}
\usepackage{color}
\usepackage[version=4]{mhchem}
\usepackage[hidelinks]{hyperref}
\usepackage{placeins}
\usepackage{mathtools}
\usepackage[normalem]{ulem}
\usepackage{lipsum}
\usepackage{array}
\usepackage{comment}
\usepackage{bm}
\usepackage{hyperref}
\usepackage{physics}
\usepackage{dsfont}
\hypersetup{colorlinks=true,citecolor=blue,linkcolor=blue,urlcolor=blue}
\usepackage{soul}
\newcommand{\bA}{\mathbf{A}}

\newcommand{\OmMax}{\overline{\Omega}}

\newcommand{\Xg}{\mathcal{X}}
\newcommand{\Hg}{\mathcal{H}}
\newcommand{\Zg}{\mathcal{Z}}

\newcommand{\CZg}{\mathrm{C}\Zg}
\newcommand{\Rg}{\mathcal{R}}
\newcommand{\CRg}{\mathrm{C}\Rg}

\usepackage{enumitem}
\setlist{nolistsep}
\begin{document}

\title{Robust Control and Entanglement of Qudits in Neutral Atom Arrays }
\author{Amir Burshtein}
\thanks{These authors contributed equally to this work.}
\affiliation{School of Physics and Astronomy, Tel Aviv University, Tel Aviv 6997801, Israel}
\author{Shachar Fraenkel}
\thanks{These authors contributed equally to this work.}
\affiliation{School of Physics and Astronomy, Tel Aviv University, Tel Aviv 6997801, Israel}
\author{Moshe Goldstein}
\affiliation{School of Physics and Astronomy, Tel Aviv University, Tel Aviv 6997801, Israel}
\author{Ran Finkelstein}
\thanks{ranf@tauex.tau.ac.il}
\affiliation{School of Physics and Astronomy, Tel Aviv University, Tel Aviv 6997801, Israel}
\affiliation{
The Center for Nanoscience \& Nanotechnology, Tel Aviv University, Tel Aviv 6997801, Israel}
\begin{abstract}
Quantum devices comprised of elementary components with more than two stable levels -- so-called qudits -- enrich the accessible Hilbert space, enabling applications ranging from fault-tolerant quantum computing to simulating complex many-body models. While several quantum platforms are built from local elements that are equipped with a rich spectrum of stable energy levels, schemes for the efficient control and entanglement of qudits are scarce. Importantly, no experimental demonstration of multi-qudit control has been achieved to date in neutral atom arrays. 
Here, we propose a general scheme for controlling and entangling qudits and perform a full analysis for the case of qutrits, encoded in ground and metastable states of alkaline earth atoms. We find an efficient implementation of single-qudit gates via the simultaneous driving of multiple transition frequencies. For entangling operations, we provide a concrete and intuitive recipe for the controlled-$\Zg$ ($\CZg$) gate for any local dimension $d$, realized through alternating single qudit and entangling pulses that simultaneously drive up to two Rydberg transitions. We further prove that two simultaneous Rydberg tones are, in general, the minimum necessary for implementing the $\CZg$ gate with a global drive. The pulses we use are optimally-controlled, smooth, and robust to realistic experimental imperfections, as we demonstrate using extensive noise simulations. This amounts to a minimal, resource-efficient, and practical protocol for realizing a universal set of gates. Our scheme for the native control of qudits in a neutral atom array provides a high-fidelity route toward qudit-based quantum computation, ready for implementation on near-term devices.
\end{abstract}

\maketitle

\section{Introduction}

Compared to qubit-based quantum platforms, systems built from higher-dimensional components, or qudits (with a local Hilbert space dimension $d>2$), offer substantial practical advantages for various aspects of quantum information processing~\cite{Wang2020}. In the context of quantum computation, these include reduced circuit depth in the implementation of algorithms~\cite{Fedorov2012,Gokhale2019} and enhanced noise resilience of protocols that establish fault tolerance (e.g., magic state distillation~\cite{Campbell2012, Campbell2014}). Qudits also support richer entanglement structures relative to those possible in qubit systems~\cite{Huber2013}. These entanglement structures are of interest both as a fundamental many-body property and as a resource for quantum metrology~\cite{RevModPhys.90.035005}. Furthermore, qudit-based quantum simulators extend our experimental reach toward various high-dimensional quantum many-body systems, such as integer-spin chains~\cite{Senko2015SpinChain}, lattice gauge theories~\cite{Zoller2022LGT,Popov2024LGT,Meth2025LGT}, continuum field theories~\cite{Calliari2025ContinuumFieldTheories}, and systems exhibiting non-abelian topological order~\cite{Tantivasadakarn2023}.

Consequently, recent years have witnessed increasing efforts to design reliable experimental platforms where native qudits can be encoded, controlled, and measured in a scalable fashion. Trapped-ion computers are arguably the most mature devices in this context, having showcased high-fidelity realization of qudit logical gates~\cite{Ringbauer2022,Hrmo2023}, and performed concrete simulation and computation tasks with native qudits~\cite{Edmunds2025,shi2025IonQuditAlgorithm}. Basic qutrit ($d=3$) and ququart ($d=4$) logic has also been successfully demonstrated using superconducting transmon circuits~\cite{Yurtalan2020SuperconductingQutrit,Blok2021QutritScrambling,Morvan2021TransmonQutrit,Goss2022}, microwave cavities~\cite{Brock2025}, and integrated photonic circuits~\cite{Chi2022}. 

\begin{figure}
    \centering
    \includegraphics[width=1\columnwidth]{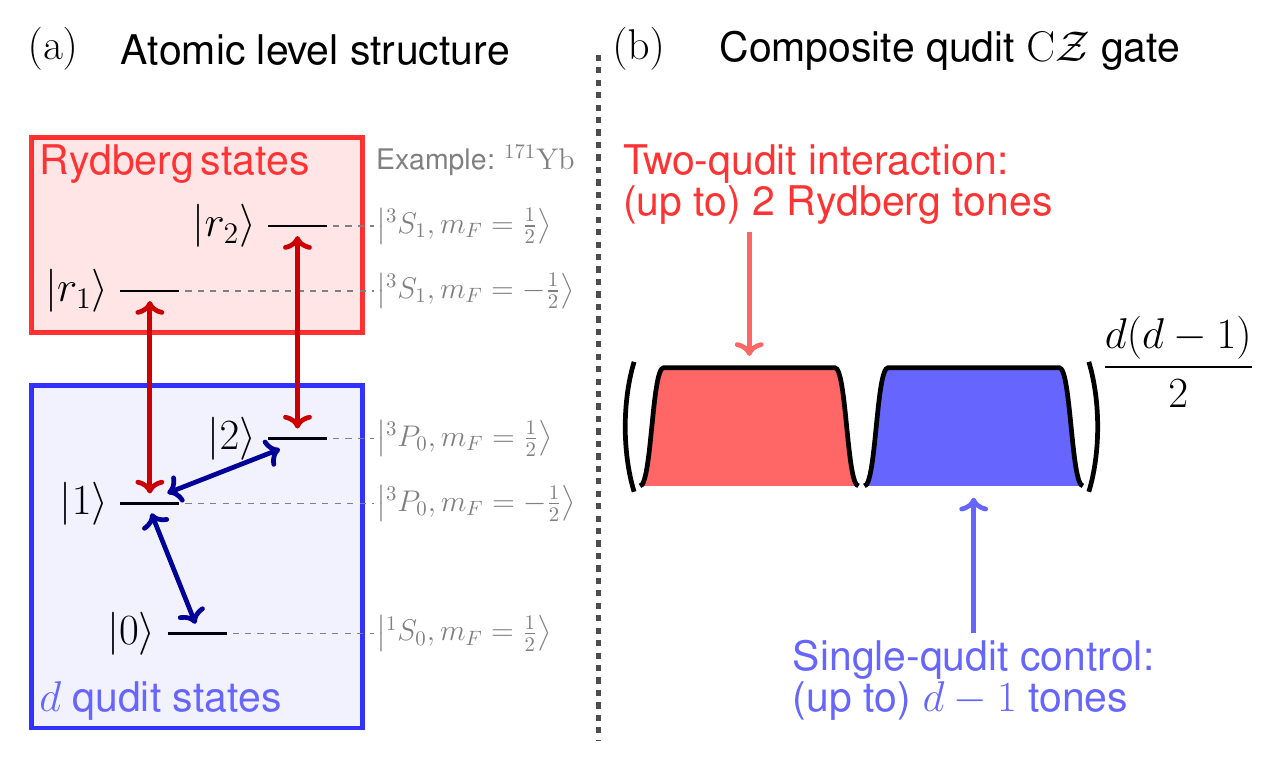}
    \caption{\textbf{Robust qudit control in neutral atoms}. \textbf{(a)} Level structure for qudit encoding, as can be found \emph{e.g.} in $^{171}{\rm Yb}$; our scheme requires $d-1$ laser tones for universal single-qudit rotations and two Rydberg tones for global entangling gates (the example shown is for $d=3$). \textbf{(b)} Robust pulse scheme for realizing the two-qudit $\CZg$ gate, which requires $\frac{d(d-1)}{2}$ entangling pulses interspersed with single-qudit rotations.}
    \label{fig:general_scheme}
\end{figure}

Neutral atom arrays -- well known for their scalability, coherence, and flexible connectivity~\cite{Bluvstein2022,Beugnon2007} -- have great appeal as potential platforms for qudit-based quantum computation. Their promise is underlined by the fact that neutral atoms naturally support multi-level encodings and enable high-fidelity entangling gates using state-selective Rydberg interactions~\cite{evered_high-fidelity_2023,Ma2023,Finkelstein2024}. First experimental attempts have demonstrated control of single qutrits in the ground-state manifold of $^{87}{\rm Rb}$ atoms ~\cite{Lindon2023RbQutrits} and all-optical control of nuclear qudits in $^{87}{\rm Sr}$~\cite{Ahmed2025} atoms.

Recently, a few theoretical proposals for implementing entangling gates in alkaline earth atomic qudits have been put forward. Ref.~\cite{Jia2024} constructed a specific scheme for encoding and controlling ququarts in $^{171}{\rm Yb}$ atoms, using their metastable and ground states. Ref.~\cite{Omanakuttan_2023_EntangleQudit} suggested a scheme for implementing general two-qudit gates for any $d$ with a single phase-modulated laser utilizing optimal control, resulting in a complex multi-parameter phase profile.

Nevertheless, a general scheme for universal multi-qudit control in neutral atoms that is simultaneously easy to calibrate, robust to noise, and applicable to any local dimension $d$ is still missing. The development of a universal control scheme that achieves these advantages would thus mark a significant milestone toward demonstrating high-fidelity computation with native qudits in neutral-atom platforms.

In this work, we propose a concrete scheme for implementing a universal gate set for qudits encoded in neutral atoms. Within our framework, ground and metastable atomic levels serve as computational qudit states with incommensurate transition frequencies that may be easily addressed individually, and Rydberg levels mediate inter-qudit entanglement (see Fig.~\ref{fig:general_scheme}(a)). The metastable nuclear spin states are efficiently coupled to the Rydberg levels via single-photon transitions, allowing for high-fidelity entangling operations.

Our control strategy combines multi-tone driving with numerical optimization techniques to achieve the efficient realization of arbitrary single-qudit gates and a class of entangling two-qudit gates. Entangling gates are implemented with global driving, that is, without need for individual single-site addressing.
This is particularly important, as pairs of neutral atoms are typically moved close to each other (with respect to the Rydberg blockade radius) in order to apply two-qudit operations, making it hard to address them individually at that time.
The two-qudit gates are constructed as pulse sequences, temporally separating fields that drive transitions between qudit states from fields that drive transitions to Rydberg states. The resulting pulses are smooth and easy to parametrize, rendering them robust to noise and implementable in current hardware.

We illustrate the power of our scheme by devising an explicit compilation of the canonical two-qudit  controlled-$\Zg$ ($\CZg$) gate built from such robust pulses, which applies to qudits of any local dimension $d$ (see Fig.~\ref{fig:general_scheme}(b)). In particular, this pulse sequence employs the simultaneous driving of two different Rydberg transitions. We prove that, for $d>3$, this two-tone Rydberg excitation corresponds to the minimal number of tones needed to implement a $\CZg$ gate with global driving.

For the case of qutrits, we perform a comprehensive analysis of elementary gates obtained using our method, deriving the precise pulse shapes that implement these gates in optimal time. We use a realistic noise model to estimate the fidelity of the qutrit $\CZg$ gate, finding a simulated fidelity of 0.994. Furthermore, we show that our method can be adapted to overcome potential crosstalk between different driving tones. Together, this amounts to a concrete recipe for universal qutrit computation that can be realized in existing neutral-atom platforms.

The remainder of the paper is organized as follows. Sect.~\ref{sec:ElementaryGates} introduces the basic gates for universal computation with qudits. In Sect.~\ref{sec:implementation}, we describe our proposed universal control scheme, including protocols for realizing general single-qudit gates and the two-qudit $\CZg$ gate. Sect.~\ref{sec:level_struct} explains how the scheme can be realistically implemented given the atomic level structures of commonly used neutral atoms. Focusing on the case of qutrits, Sect.~\ref{sec:Imperfections} presents noise simulations that demonstrate the robustness of our scheme to realistic implementation imperfections, and discusses advantages over qubit-based schemes. Finally, Sect.~\ref{sec:summary} summarizes our findings and presents several directions for future studies.
Technical details are relegated to the Appendixes.

\section{Elementary operations for universal qudit-based computation}\label{sec:ElementaryGates}

To facilitate universal unitary control over multiple qudits with local dimension $d$, it is sufficient to be able to apply any single-qudit unitary operator in ${\rm SU}(d)$ (i.e., any $d\times d$ unitary up to a global phase) in addition to a certain two-qudit entangling unitary gate. A multi-qudit unitary can always be represented as a sequence of such gates~\cite{PhysRevA.62.052309,Brylinski2002}. This requirement will guide us in designing our computation scheme.

Let us present several elementary unitary gates that will be central to our subsequent analysis. We define generalized Pauli gates through their action on the qudit computational basis states $\ket{j}$ ($j=0,1,\ldots,d-1$):
\begin{equation}\label{eq:PauliDef}
    \Xg|j\rangle=|j+1({\rm mod}\,d)\rangle,\quad\Zg|j\rangle=\omega^j|j\rangle,
\end{equation}
where $\omega=\exp\!\left(i2\pi/d\right)$. A controlled-$\Zg$ operation, which is a two-qudit entangling gate, is defined as
\begin{equation}\label{eq:CZDef}
\CZg|j,k\rangle=\omega^{jk}|j,k\rangle.
\end{equation}
In addition, we may define the Hadamard gate $\Hg$ and a set of gates $\Rg_k(\theta)$ assigning an arbitrary phase $\theta$ only to the computational basis state $|k\rangle$:
\begin{equation}\label{eq:CliffordGeneratorsDef}
    \Hg|j\rangle=\frac{1}{\sqrt{d}}\sum_{k=0}^{d-1}\omega^{jk}|k\rangle,\quad\Rg_k(\theta)|j\rangle=e^{i\theta\delta_{jk}}|j\rangle.
\end{equation}

These elementary gates assume an especially important role when $d$ is prime. Indeed, for prime $d$, it becomes possible to generate all unitary operations using only the gates $\Hg$, $\CZg$, and two diagonal single-qudit gates of the form $\prod_{k=0}^{d-1}\Rg_k(\theta_k)$~\cite{Howard2012_QuditGates}. Alternatively, if we remove one of the diagonal single-qudit gates, we can still generate the restricted set of Clifford operators~\cite{Clark_2006}. These are useful for the magic-state model of fault-tolerant quantum computation, which relies on the system evolving only through Clifford operations from appropriate initial states~\cite{Bravyi2005,Campbell2012}. %For completeness, we provide more details on these constructions in Appendix~\ref{app:UniGateSet}.
In any case, the existence of such a small universal set of gates is an extremely powerful simplification, and ensuring the robust realization of these particular gates paves the way toward a viable computation platform. 

\section{Scheme for universal multi-qudit control}\label{sec:implementation}
In this section, we introduce our general scheme for universal quantum computation with qudits encoded in neutral atoms. Our scheme applies to qudits of any local dimension $d$, and relies on multi-frequency control. It requires $d-1$ laser tones for single-qudit operations, and at least one laser tone that drives transitions to the Rydberg manifold to entangle qudits.

We focus on the set of elementary gates introduced in Sect.~\ref{sec:ElementaryGates} and design simple pulses that realize these gates in optimal time (which is the simplest and most important criterion, given that the dominant errors, especially Rydberg state decay, are incoherent, and are thus are typically minimized together with the overall gate duration). The optimization procedure utilizes established techniques for quantum optimal control~\cite{Merkel_2008_OptimalControl,Smith_2013_OptimalControl,Anderson_2015_OptimalControl,Lysne_2020_OptimalControl,Omanakuttan_2021_OptimalControl,Jandura2022,Omanakuttan_2023_EntangleQudit}, and specifically the Gradient Ascent Pulse Engineering (GRAPE) method~\cite{KHANEJA2004}. We provide the details of our optimization method in Appendix~\ref{app:GRAPE}. Concretely, we show the resulting pulses for the case of qutrit gates. Remarkably, the optimization results in pulses with fixed amplitudes and smooth phase gradients, which are readily implementable in current experiments thanks to their easy parametrization. Thus, by combining the versatility of the multi-tone setup with pulse optimization, we are able to fully control the qudits through pulses that are both efficient and robust to noise and realistic imperfections.

\subsection{Universal single-qudit control}\label{subsec:SingleQuditGate}
Here we present our scheme for applying any ${\rm SU}(d)$ operator to a single qudit using $d-1$ independent lasers or laser tones. Each of these fields is tuned near the transition frequency $\omega_{j_1,j_2}$ between a pair of qudit levels, $|j_1\rangle$ and $|j_2\rangle$. The corresponding Hamiltonian (in the interaction picture, and following a rotating wave approximation) is given by
\begin{equation}\label{eq:RabiHamiltonian}
    H_{j_1,j_2}=\frac{\Omega_{j_1,j_2}(t)}{2}|j_1\rangle\langle j_2| + {\rm h.c.},
\end{equation}
and we assume that we can modulate the amplitude and phase of $\Omega_{j_1,j_2}$, and that the amplitude (Rabi frequency) is upper-bounded by a maximal possible value $\OmMax$.

We require the $d-1$ transitions resonantly driven by the lasers to have incommensurate frequencies $\omega_{j_1,j_2}$, and to connect all of the qudit levels. For the latter condition to be satisfied, we may assume that the lasers drive the transitions $|j\rangle\leftrightarrow|j+1\rangle$ for $j=0,1,\ldots,d-2$, though any other permutation of the qudit levels will work similarly well. The evolution of the qudit is then governed by the Hamiltonian $\sum_{j=0}^{d-2} H_{j,j+1}$, and the tones $\Omega_{j,j+1}(t)$ can be determined according to the particular evolution we wish to realize. This is sufficient for realizing any operator in ${\rm SU}(d)$, as we prove in Appendix~\ref{app:SingleQuditControl}. For any particular operator $U\in{\rm SU}(d)$, we choose the tones that realize $U$ in optimal time, which we find using the optimization method detailed in Appendix~\ref{app:GRAPE}.

\begin{figure}
    \centering
    \includegraphics[width=1\linewidth]{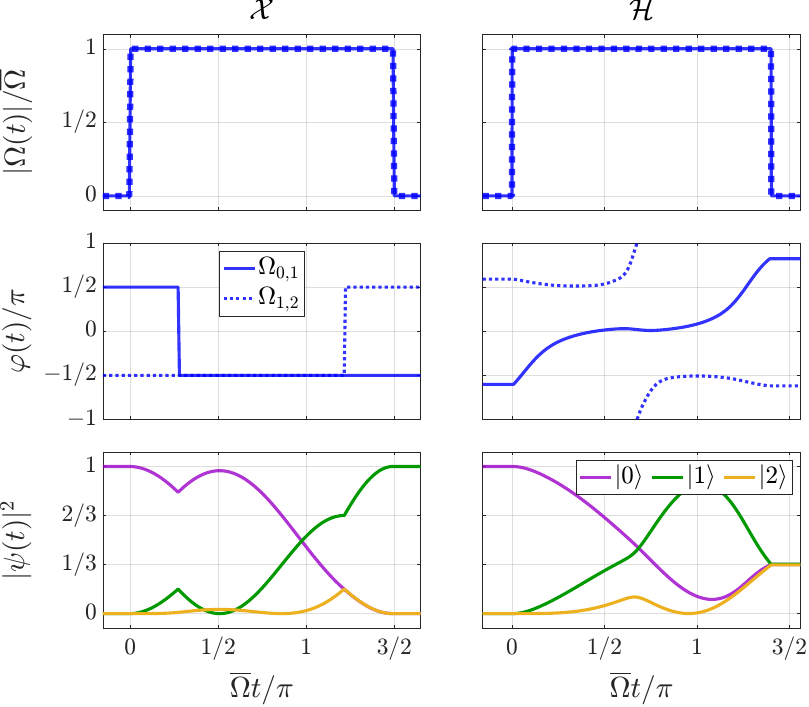}
    \caption{\textbf{Single-qutrit gates.} Optimal-time pulse shapes implementing the $\Xg$ and $\Hg$ single-qutrit gates. The lasers operate at maximal intensity $\OmMax$ (top row), and the gates are implemented by the modulation of the phase (middle row). Note that the apparent discontinuity of the phase of $\Omega_{1,2}$ in $\Hg$ is simply a $2\pi$ phase wrap. The bottom row shows the population evolution, starting from $\psi(0)=\ket{0}$.}
    \label{fig:single_qudit}
\end{figure}

To test the effectiveness of our method, we explicitly apply it to the case of qutrits ($d=3$), assuming two lasers with equal Rabi frequency $\OmMax$ that address the transitions $|0\rangle\leftrightarrow|1\rangle$ and $|1\rangle\leftrightarrow|2\rangle$. In Fig.~\ref{fig:single_qudit}, we show the optimal-time pulse shapes that realize the gates $\Xg$ and $\Hg$ [defined in Eqs.~\eqref{eq:PauliDef} and \eqref{eq:CliffordGeneratorsDef}, respectively], which turn out to have a fixed maximal amplitude. Interestingly, we find that the $\Xg$ gate is realized using a single pulse divided into three sections with alternating $0$ and $\pi$ phase differences between the fields $\Omega_{0,1}$ and $\Omega_{1,2}$. Note that the simultaneous operation of the two laser fields shortens the total gate duration: The same gate may be implemented using consecutive $\pi$ pulses of the two fields, which would result in a total duration of $T = 2\pi/\OmMax$, compared with the optimal $T \approx 3\pi/(2\OmMax)$ shown in Fig.~\ref{fig:single_qudit}.

Let us also note that because our gate scheme relies on lasers that induce transitions between qudit levels, using it to realize diagonal phase gates such as $\Zg$ and $\Rg_k(\theta)$ [Eqs.~\eqref{eq:PauliDef} and \eqref{eq:CliffordGeneratorsDef}, respectively] is rather inefficient. It is favorable to implement such diagonal gates as virtual zero-duration gates -- that is, to embed them in subsequent operations in the circuit through appropriate phase offsets of the Rabi drives~\cite{McKay2017VirtualGates}.

\begin{figure*}
    \centering
    \includegraphics[width=1\linewidth]{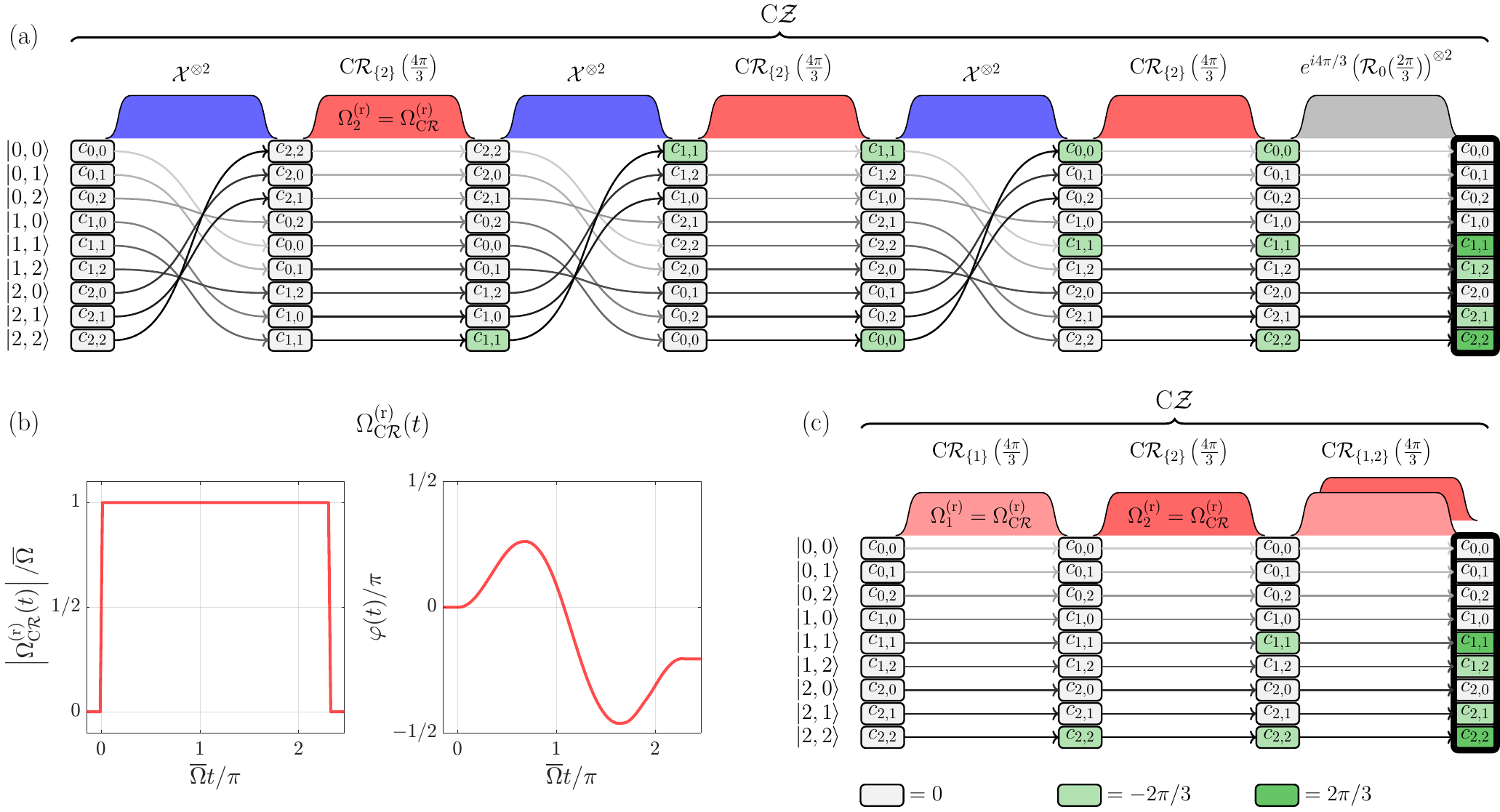}
    \caption{\textbf{Composite qutrit $\CZg$ gate.} \textbf{(a)} Implementation using a single Rydberg state (i.e., only $\Omega_2^{(\mathrm{r})}$) and single-qudit permutations, following Eq.~\eqref{eq:CZ_qutrit_single_ryd}, showing the evolution of the input state under the applied pulses.
    \textbf{(b)} Amplitude and phase of the optimal-time pulse shape $\Omega_{\CRg}^{(\mathrm{r})}$ implementing the $\CRg\left(\frac{4\pi}{3}\right)$ gates.
    \textbf{(c)} Implementation using two available Rydberg states, following Eq.~\eqref{eq:CZ_general_qutrit}. The first pulse has $\Omega_1^{(\mathrm{r})} = \Omega_{\CRg}^{(\mathrm{r})}$ and $\Omega_2^{(\mathrm{r})} = 0$, the second pulse has $\Omega_1^{(\mathrm{r})} = 0$ and $\Omega_2^{(\mathrm{r})} = \Omega_{\CRg}^{(\mathrm{r})}$, and the third pulse activates both lasers simultaneously.}
    \label{fig:cz3_pulsed}
\end{figure*}

\subsection{Entangling two-qudit gate}\label{subsec:CZ_Gate}

We now turn to providing a general protocol for implementing an entangling two-qudit gate, which is necessary and sufficient for universal control (see Sect.~\ref{sec:ElementaryGates}). Specifically, we focus on the canonical $\CZg$ gate [defined in Eq.~\eqref{eq:CZDef}] and construct it from a sequence of simple global pulses that enable its robust realization. We emphasize that our strategy can be applied to construct any symmetric two-qudit phase gate for any $d$.

The basic ingredient of our protocol relies on the standard implementation of the global controlled-phase gate designed for neutral atom qubits~\cite{Levine2019,Jandura2022}. It involves one Rydberg level $\ket{r}$ in each atom, and requires a laser field $\Omega^{({\mathrm r})}(t)$ that resonantly drives the transition between $\ket{r}$ and one of the qudit levels $\ket{j}$. We assume a perfect Rydberg blockade (infinitely strong interaction) between the Rydberg levels of the two atoms; the effect of imperfect blockade can be compensated through minor calibration~\cite{evered_high-fidelity_2023}.

By properly choosing the pulse shape of the field $\Omega^{({\mathrm r})}(t)$, it is then possible to impart an arbitrary phase factor $e^{i\theta}$ to the two-qudit state $\ket{j,j}$, without changing any other two-qudit state~\footnote{A pulse of this type actually imparts a phase $e^{i(2\chi+\theta)}$ to the state $\ket{j,j}$ and a phase $e^{i\chi}$ to any state of the form $\ket{j,k}$ or $\ket{k,j}$ with $k\neq j$, where $\chi$ depends on the specific pulse shape. However, up to single-qudit phase gates, this operation is equivalent to the same operation with $\chi=0$, so we ignore this detail henceforth~\cite{Levine2019}.}. We let $\CRg_{\{j\}}(\theta)$ denote this operation, so that
\begin{equation}
    \CRg_{\{j\}}(\theta)\ket{k_1,k_2}=e^{i\theta\delta_{j,k_1}\delta_{j,k_2}}\ket{k_1,k_2}.
\end{equation}
This operator generally entangles the two qudits, as long as $\theta$ is not an integer multiple of $2\pi$. In fact, for qubits this is already the $\CZg$ gate if we set $j=1,\theta=\pi$.
As in the case of single-qudit gates, there are various pulse shapes of $\Omega^{({\mathrm r})}(t)$ that implement the same operator $\CRg_{\{j\}}(\theta)$. Using the optimization method detailed in Appendix~\ref{app:GRAPE}, we find the pulse that implements the gate in optimal time~\cite{Jandura2022}.

Given that $\CRg_{\{j\}}(\theta)$ is an entangling gate, in principle it can be used together with single-qudit gates to construct any multi-qudit gate, including the $\CZg$ gate. However, this does not guarantee that such a construction will be efficient. We now present a simple and robust protocol for constructing the $\CZg$ gate for any $d$, which involves only global pulses and generally requires two distinct Rydberg levels to be coupled to the qudit levels. We will begin by explaining the particular example of qutrits, and then generalize. 

\subsubsection{Constructing the qutrit $\CZg$ gate}
Here we present two possible constructions of the qutrit $\CZg$ gate. In the first, we show that the two-qutrit $\CZg$ gate can be implemented as a pulse sequence that includes $\CRg_{\{2\}}\!\left(\frac{4\pi}{3}\right)$ together with single-qudit phase gates and $\Xg$ gates. Indeed, up to an irrelevant global phase, it may be written as follows:
\begin{equation}\label{eq:CZ_qutrit_single_ryd}
    \CZg = \left(\Rg_0\!\left(\frac{2\pi}{3}\right)\right)^{\otimes2}\left[\CRg_{\{2\}}\!\!\left(\frac{4\pi}{3}\right)\Xg^{\otimes2}\right]^3.
\end{equation}
The different steps in this gate sequence are schematically broken down in Fig.~\ref{fig:cz3_pulsed}(a). The idea behind this protocol is that, while $\CRg_{\{2\}}\!\left(\frac{4\pi}{3}\right)$ imparts a nontrivial phase only to the $\ket{2,2}$ state, it may be used to impart the same phase also to the states $\ket{0,0}$ and $\ket{1,1}$ if we use the $\Xg$ gate to permute the qudit levels. In this manner, the levels $\ket{0}$ and $\ket{1}$ do not need to be coupled directly to any Rydberg level.

The shape of the optimal-time pulse realizing $\CRg_{\{2\}}\!\left(\frac{4\pi}{3}\right)$ is plotted in Fig.~\ref{fig:cz3_pulsed}(b). As in the case of the single-qutrit gates we examined, this is a fixed-amplitude pulse with a smoothly varying phase. Note that the pulse we obtain is qualitatively similar to the two-qubit $\CZg$ gate~\cite{Jandura2022,evered_high-fidelity_2023} but with a shorter duration. Recall that for the $\Xg$ gate appearing in Eq.~\eqref{eq:CZ_qutrit_single_ryd}, the optimal-time pulse was already given in Fig.~\ref{fig:single_qudit}.

A second, more efficient, qutrit $\CZg$ gate protocol becomes available if we allow excitations to two distinct Rydberg levels. Namely, let us assume that each atom has two Rydberg levels $\ket{r_1}$ and $\ket{r_2}$, and let $\Omega_1^{({\mathrm r})}$ and $\Omega_2^{({\mathrm r})}$ denote two laser fields that resonantly drive the transitions $\ket{1}\leftrightarrow\ket{r_1}$ and $\ket{2}\leftrightarrow\ket{r_2}$, respectively. We assume that the two transitions can be driven independently (e.g., using incommensurate frequencies or cross-polarized beams, as discussed in Sect.~\ref{sec:level_struct}; the possible relaxation of this assumption is discussed in Sect.~\ref{sec:Imperfections}). Furthermore, we assume perfect Rydberg blockade between the two different Rydberg levels $\ket{r_1}$ and $\ket{r_2}$~\cite{Jia2024}.

Turning on the field $\Omega_1^{({\mathrm r})}$ while keeping $\Omega_2^{({\mathrm r})}$ off applies the operator $\CRg_{\{1\}}\!\left(\theta\right)$; the opposite choice will, of course, realize $\CRg_{\{2\}}\!\left(\theta\right)$. We can also simultaneously activate the two fields with the same pulse shapes that realize $\CRg_{\{1\}}\!\left(\theta\right)$ and $\CRg_{\{2\}}\!\left(\theta\right)$ when they are activated alone. This realizes the gate $\CRg_{\{1,2\}}\!\left(\theta\right)$, where we define
\begin{equation}\label{eq:ConditionalRDef2}
    \CRg_{\{j,m\}}(\theta)\ket{k_1,k_2}=\begin{cases}
        e^{i\theta}\ket{k_1,k_2} & k_1,k_2\in\{j,m\},\\
        \ket{k_1,k_2} & {\rm otherwise}.
    \end{cases}
\end{equation}
That is, $\CRg_{\{1,2\}}\!\left(\theta\right)$ imparts the phase factor $e^{i\theta}$ to all the states $\ket{1,1}$, $\ket{1,2}$, $\ket{2,1}$, and $\ket{2,2}$. Notably, a global field that drives only a single Rydberg transition cannot impart a non-trivial phase to a state $\ket{j,m}$ with $j\neq m$, which points to the advantage of driving the two transitions together.

We observe that we can construct the $\CZg$ gate through the following pulse sequence [illustrated in Fig.~\ref{fig:cz3_pulsed}(c)]:
\begin{equation}\label{eq:CZ_general_qutrit}
    \CZg = \CRg_{\{1,2\}}\!\left(\frac{4\pi}{3}\right)\CRg_{\{2\}}\!\left(\frac{4\pi}{3}\right)\CRg_{\{1\}}\!\left(\frac{4\pi}{3}\right).
\end{equation}
The duration of this pulse sequence is shorter compared with that of the pulse sequence in Eq.~\eqref{eq:CZ_qutrit_single_ryd}. It requires three pulses equivalent to $\CRg_{\{j\}}\!\left(\frac{4\pi}{3}\right)$, while Eq.~\eqref{eq:CZ_qutrit_single_ryd} requires three additional steps of single-qudit rotations. 

Crucially, as we will soon show, the $\CRg_{\{1,2\}}\!\left(\frac{4\pi}{3}\right)$ gate does not only shorten the total duration of the composite qutrit $\CZg$ gate, but also sets the basis to generalize the protocol to any dimension $d$. In fact, we prove below that the simultaneous driving of two Rydberg transitions is needed to implement the $\CZg$ gate for $d>3$ with global drives.

In principle, the $\CZg$ gate may also be implemented by simultaneously driving both intra-qudit and Rydberg transitions, using numerical optimization to find the appropriate amplitude and phase profiles for the different driving fields. This is discussed in Appendix~\ref{app:CZ_fourier}. While this approach may allow one to shorten to overall gate duration, it suffers from two crucial drawbacks. First of all, it does not provide a recipe that can be generalized to higher qudit dimensions, as opposed to the general scheme we provide below. Furthermore, compared with the pulsed version of Eq.~\eqref{eq:CZ_general_qutrit}, this approach is considerably more challenging to calibrate, which would be necessary in the presence of experimental error sources, as discussed below in Sect.~\ref{sec:Imperfections}. Any modification, either to increase the qudit dimension or to account for experimental error sources, would require the re-optimization of the laser profiles, with computation times that grow substantially with the qudit dimension due to the increase in the number of degrees of freedom.

\subsubsection{No-go theorem for $d>3$ using a single Rydberg transition}
In our construction of the qutrit $\CZg$ gate, the source of its simplicity and robustness is its composite pulse structure, i.e., the temporal separation of single-qudit rotations and Rydberg entangling pulses. To generalize this construction to any $d$, we first identify a fundamental limitation: For $d>3$, the $\CZg$ gate cannot be realized as a composite pulse sequence with a global drive that addresses only a single Rydberg transition. This is possible only for $d\le3$, as Eq.~\eqref{eq:CZ_qutrit_single_ryd} showed.

This limitation is established by the following theorem. Consider a two-qudit phase gate $U$ (i.e., a unitary operator that is diagonal in the computational basis) that is of the following general form:
\begin{equation}\label{eq:OneRydPhaseGate}
    U={W_n}^{\otimes2}\CRg_{\{2\}}(\theta_n){W_{n-1}}^{\otimes2}\cdot\ldots\cdot\CRg_{\{2\}}(\theta_1){W_0}^{\otimes 2},
\end{equation}
where $W_k$ are single-qudit unitaries, and $\theta_k$ are controlled phases. Then, $U$ must act on the states $\ket{j,m}$ with $j\neq m$ as 
\begin{equation}\label{eq:PhaseOpNoGo}
    U\ket{j,m}=e^{i(\xi_j+\xi_m)}\ket{j,m}\quad\text{(if $j\neq m$)}.
\end{equation}
We prove this theorem in Appendix~\ref{app:CZ_no_go}. The form of $U$ in Eq.~\eqref{eq:OneRydPhaseGate} stems from the global driving (so that single-qudit operations are symmetric) and the coupling of only one qudit level to the Rydberg manifold (denoted here arbitrarily as $\ket{2}$). Note that the general form of Eq.~\eqref{eq:OneRydPhaseGate} includes the qutrit $\CZg$ gate in Eq.~\eqref{eq:CZ_qutrit_single_ryd}.

Eq.~\eqref{eq:PhaseOpNoGo} represents a strict limitation on the two-qudit phase gates that can be implemented using our optimized, resource-efficient scheme. In particular, it prohibits the realization of the $\CZg$ gate for $d>3$. To see this, assume for the sake of contradiction that $U=\CZg$. Then since $U\ket{0,j}=\ket{0,j}$ for any $j>0$, we must also have that $\xi_j=-\xi_0$ for any $j>0$. So $U\ket{j,m}=e^{-i2\xi_0}\ket{j,m}$ for any $m>j>0$, independently of $j,m$. This cannot correspond to the desired action of $\CZg$ for $d>3$.

% Thus, a global realization of $\CZg$ in the form of a pulse sequence that separates single-qudit tones from Rydberg tones requires at least two different Rydberg tones that can be applied simultaneously.

% For general $d$, the addressing of two distinct transitions to the Rydberg manifold becomes a minimal requirement 

\subsubsection{Constructing the general qudit $\CZg$ gate}
We have concluded that, for $d>3$, the $\CZg$ gate cannot be implemented as a composite pulse sequence using global drives if only one qudit level is directly coupled to a Rydberg level. In contrast, we now show that the simultaneous driving of two Rydberg transitions allows us to straightforwardly generalize the $\CZg$ protocol of Eq.~\eqref{eq:CZ_general_qutrit} to any $d$. This general protocol thus requires the minimal number of distinct transitions to the Rydberg manifold.

The general protocol is dictated by the following decomposition of the $\CZg$ gate:
\begin{equation}\label{eq:CZ_general_qudit}
    \CZg = \prod_{j=1}^{d-1}\CRg_{\{j\}}\!\left(\theta_{j,j}\right)\prod_{1\le j<m\le d-1}\CRg_{\{j,m\}}\!\left(\theta_{j,m}\right),
\end{equation}
where we have defined the phases
\begin{equation}\label{eq:PhaseMatrix}
    \theta_{j,m}=\frac{2\pi}{d}jm+\left(\frac{2\pi}{d}j^2-\pi j(d-1)\right)\delta_{jm}.
\end{equation}
The sequence in Eq.~\eqref{eq:CZ_general_qudit} includes $d-1$ pulses that employ a single driving tone and $\frac{1}{2}(d-1)(d-2)$ pulses that employ two simultaneous driving tones. Naively, this protocol requires $d-1$ Rydberg levels $\ket{r_j}$ and $d-1$ independent laser fields that drive the transitions $\ket{j}\leftrightarrow\ket{r_j}$ for $j=1,\ldots,d-1$. However, only two Rydberg levels and two qudit states coupled to them are necessary in practice, as the entangling pulses can be interspersed with single-qudit gates that permute the qudits states, similarly to how we used the $\Xg$ gates in Eq.~\eqref{eq:CZ_qutrit_single_ryd}.

If more than two Rydberg levels can be coupled to different qudit levels, then the protocol requires fewer permutation gates and becomes more efficient. This is ultimately limited by the number of independent Rydberg transitions that we can identify in the atomic level structure. If such additional Rydberg levels are indeed available, they also open up the possibility of simultaneously driving more than two Rydberg transitions, potentially further expediting the protocol. For instance, in the case of ququints ($d=5$) we found that by allowing the simultaneous driving of three Rydberg transitions, we can reduce the number of entangling pulses from 10 [as in Eq.~\eqref{eq:CZ_general_qudit}] to 7.

\section{Proposed experimental implementation}\label{sec:level_struct}
% Yb choice
Our proposed protocol is general and may be implemented with different types of neutral atoms or extended to other platforms. Here, for concreteness, we consider an experimental implementation with arrays of alkaline-earth neutral atoms, which include multiple stable levels spanned by the nuclear spin orientation in both the ground and metastable electronic levels. We note that such qudit encoding can also be realized in the hyperfine structure of the ground state in alkali atoms.
As a concrete example, we analyze the case of qutrits encoded in $^{171}{\rm Yb}$, illustrated in Fig.~\ref{fig:general_scheme}(a). This specific species was recently used to implement high-fidelity entangling gates using qubits encoded in the metastable electronic level~\cite{Ma2023}, as well as high-fidelity control and readout of optical qubits and ground-state qubits~\cite{Lis2023}.
As a minimal extension, we suggest adding a single electronic ground-state level. Under a moderate magnetic field, the ground and metastable nuclear spin manifolds exhibit different splittings because of different Land\'e $g$-factors. Single qudit rotations can be realized by combining a narrow-line clock laser (acting on the optical transition) and a Raman laser (acting on the metastable manifold). Measurements can be performed through a combination of qudit rotations and spin-resolved detection~\cite{Senoo2025}.

To realize multi-tone Rydberg driving in such systems, one can either utilize a single frequency-modulated beam with identical polarizations across tones, or combine cross-polarized beams. The first configuration offers the advantage of a single beam and a single modulator; however, it may induce off-resonant crosstalk between the different Rydberg transitions. Using circular cross-polarized beams ($\sigma^+$ and $\sigma^-$) to address the different Rydberg levels can minimize such crosstalk at the expense of a slightly more complicated setup. As an example, in the case of $^{171}{\rm Yb}$ one can utilize the $F=1/2$ Rydberg manifold~\cite{peper_spectroscopy_2025} and address the $m_F=1/2\rightarrow m_F'=-1/2$ and $m_F=-1/2\rightarrow m_F'=1/2$ transitions (see Appendix~\ref{app:crosstalk}) with circular cross-polarized fields. Note that the interaction energy between atoms in different Zeeman sub-levels of the same Rydberg level is identical to the interaction energy between atoms in the same Zeeman sub-level to a good approximation for high-lying Rydberg states~\cite{Jia2024}. Thus, our proposed qutrit gates may be implemented in current hardware with no need for significant experimental upgrades. 

\section{Noise and comparison with qubit-based implementations}\label{sec:Imperfections}
The protocols proposed in Sect.~\ref{sec:implementation} offer an intuitive and scalable recipe to achieve universal control. We now show that our gates achieve high fidelities in the presence of realistic noise sources and imperfections in present-day hardware. We model and simulate four dominant error sources~\cite{scholl_erasure_2023,Tsai2025,peper_spectroscopy_2025}: (i) Shot-to-shot detuning fluctuations (due to Doppler shifts and laser frequency noise), (ii) shot-to-shot laser intensity noise, (iii) Rydberg decay, and (iv) crosstalk between the Rydberg drive tones. As mentioned above, using circularly-polarized light for the Rydberg tones could eliminate (iv) as an error source. In this section, we simulate the qutrit $\CZg$ gate in the absence of Rydberg tone crosstalk. In Appendix~\ref{app:crosstalk}, we show how our protocol could be adjusted when the crosstalk cannot be avoided, and simulate the qutrit $\CZg$ gate in its presence. We benchmark the gate with a Monte Carlo wavefunction simulation using the quantum jump method \cite{Molmer93}, testing the gate fidelity by applying it to the equal-superposition state $\frac{1}{3}\left(\ket{0}+\ket{1}+\ket{2}\right)^{\otimes 2}$. In each scenario discussed below, we average the fidelity over $10^6$ simulated trajectories. All fidelities reported in this section are obtained within statistical errors of less than $0.001$.

\subsection{Simulation in the absence of Rydberg tone crosstalk}
As discussed in Sect.~\ref{sec:level_struct}, in the qutrit setup, using cross-polarized laser beams eliminates the crosstalk between different Rydberg transitions. In that case, these transitions may be driven independently, yielding optimal performance of the protocol proposed in Eq.~\eqref{eq:CZ_general_qutrit} and depicted in Fig.~\ref{fig:cz3_pulsed}(c). In our simulation, we use a Rabi frequency of $\OmMax=2\pi\times 5~\textrm{MHz}$ and a Rydberg lifetime of $60~\mu\textrm{s}$. To simplify the quantum jump simulation, we assume that spontaneous decay from the Rydberg states always occurs to the ground state $\ket{0}$. Note that spontaneous decay to any level should destroy the gate operation fidelity in a similar fashion and lead to a random output state with small overlap with the target state. Furthermore, the branching ratio in Yb atoms (which we consider as the proposed experimental realization in Sect.~\ref{sec:level_struct}) renders the decay probabilities into $\ket{1}$,$\ket{2}$ very small~\cite{Ma2023}. We also use a finite blockade strength $V=2\pi\times 220~\textrm{MHz}$, which we assume to be uniform across all Rydberg levels~\cite{Jia2024}. The shot-to-shot detuning is sampled from a normal distribution with a standard deviation of $40~\textrm{KHz}$, and the laser intensity noise with a relative variance of 0.008.

To implement the $\CRg$ gates, we modify the pulse shape $\Omega_{\CRg}^{(\mathrm{r})}(t)$ shown in Fig.~\ref{fig:cz3_pulsed}(b) and re-optimize the Rydberg driving pulse with finite amplitude rise and fall times, to avoid abrupt on-off laser switching. We note that such adjustments in the presence of systematic experimental imperfections (such as finite rise and fall times, or finite blockade strength) may be performed easily, since $\Omega_{\CRg}^{(\mathrm{r})}(t)$ is a smooth function that can be faithfully represented by a few parameters. One may thus write a simple ansatz for the pulse shape implementing $\CRg_S(\theta)$ for any $\theta$ and any set $S$ of target states, and use the ansatz to calibrate the pulses~\cite{evered_high-fidelity_2023}.

We find that the qutrit $\CZg$ gate is implemented with a fidelity of 0.994. We further test the fidelities of the $\CRg$ gates comprising the gate sequence in Eq.~\eqref{eq:CZ_general_qutrit}; we find that $\CRg_{\{1\}}\!\left(\frac{4\pi}{3}\right)$, $\CRg_{\{2\}}\!\left(\frac{4\pi}{3}\right)$, and $\CRg_{\{1,2\}}\!\left(\frac{4\pi}{3}\right)$ are implemented with fidelities 0.998, 0.998, and 0.997, respectively. Note that the degraded fidelity of $\CRg_{\{1,2\}}\!\left(\frac{4\pi}{3}\right)$ compared with $\CRg_{\{1\}}\!\left(\frac{4\pi}{3}\right)$ and $\CRg_{\{2\}}\!\left(\frac{4\pi}{3}\right)$ is expected, due to the doubled population in the Rydberg levels, which is the main error source. In Appendix~\ref{app:crosstalk}, we show that the effect of crosstalk, if it cannot be avoided, can be counteracted by simple modifications to the $\CRg$ pulses, resulting in a qutrit $\CZg$ fidelity of 0.993 that is close to the fidelity in the absence of crosstalk.

% The fidelities are summarized in Table \ref{tab:cz_fids}; see Appendix \ref{app:crosstalk} for details on the simulation in the presence of Rydberg tone crosstalk.

\begin{comment}
    \begin{table}[t]
    \centering
    \begin{tabular}{|c|c|c|}
        \hline
         & Without crosstalk & With crosstalk \\
         \hline \hline
        $\bm{\CZg}$ & $\bm{0.994(1)}$ & $\bm{0.993(1)}$ \\ \hline
        $\CRg_{\{1\}}\left(\frac{4\pi}{3}\right)$ & 0.998(1) & 0.998(1) \\ \hline
        $\CRg_{\{2\}}\left(\frac{4\pi}{3}\right)$ & 0.998(1) & 0.998(1) \\ \hline
        $\CRg_{\{1,2\}}\left(\frac{4\pi}{3}\right)$ & 0.997(1) & 0.997(1) \\ \hline
    \end{tabular}
    \caption{Simulated fidelities of the qutrit $\CZg$ and $\CRg$ gates, without and with crosstalk between the Rydberg tones.}
    \label{tab:cz_fids}
\end{table}
\end{comment}

In the special case of qutrits, one may also examine the protocol in Eq.~\eqref{eq:CZ_qutrit_single_ryd}, which utilizes a single Rydberg tone. This protocol involves three applications of $\CRg_{\{2\}}\!\left(\frac{4\pi}{3}\right)$, such that the combined fidelity of the entangling operations should be approximately given by $F_{\CRg_{\{2\}}(\frac{4\pi}{3})}^3\approx 0.996$. The choice between the protocols of Eqs.~\eqref{eq:CZ_qutrit_single_ryd} and~\eqref{eq:CZ_general_qutrit} then boils down to the fidelity of the single-qutrit rotations.

\subsection{Predicted scaling of $\CZg$ fidelity with $d$}
\begin{figure}
    \centering
    \includegraphics[width=1\columnwidth]{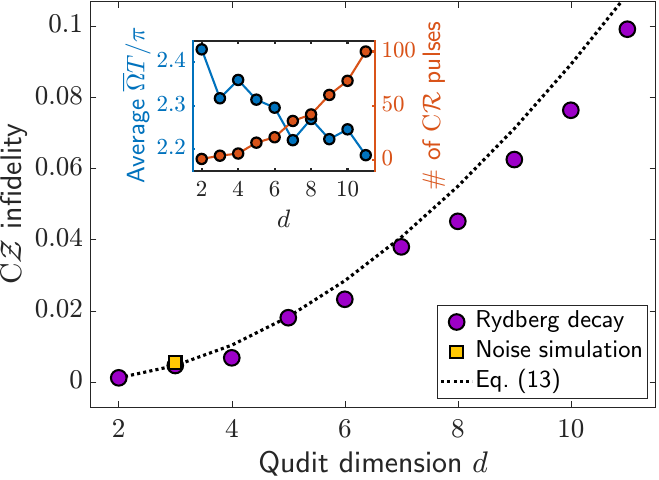}
    \caption{\textbf{Predicted infidelity for the qudit $\CZg$ gate from Rydberg decay.} We assume a Rydberg lifetime of $60~\mu\textrm{sec}$ and a Rabi frequency of $\OmMax=2\pi\times 5~\textrm{MHz}$. The dotted black line shows the predicted scaling of the infidelity with the qudit dimension, given by Eq.~\eqref{eq:infid_scaling}. The yellow square marks the infidelity of the qutrit $\CZg$ gate obtained from the noise simulation, accounting also for shot-to-shot noise and a finite Rydberg blockade. % The bottom inset shows the pulse durations of $\CRg(\theta)$ as a function of $\theta$.
    % The inset shows the predicted infidelity scaling of the ququint $\CZg$ gate with the Rabi frequency, assuming again a Rydberg lifetime of $60~\mu\textrm{sec}$.
    The inset shows the average $\CRg$ pulse duration for each $d$ and the weighted number of non-trivial $\CRg$ pulses (i.e., with $\theta \neq 0$), where each two-tone pulse is counted as two pulses.}
    \label{fig:infid_scaling}
\end{figure}

The dominance of Rydberg decay over other sources of infidelity allows us to estimate how the $\CZg$ fidelity scales with the qudit dimension $d$. We evaluate the average population in the Rydberg states, $\chi_{\textrm{ryd}}$, during the time evolution under the $\CRg$ pulses in Eq.~\eqref{eq:CZ_general_qudit} (see Appendix~\ref{app:Rydberg_Time} for details). The Rydberg decay infidelity of each pulse is approximately given by $1 - e^{-\chi_{\textrm{ryd}}T/\tau_{\textrm{ryd}}}$, where $T$ is the pulse duration and $\tau_{\textrm{ryd}}$ is the Rydberg lifetime. From the product of these exponentials, one may estimate the infidelity of the $\CZg$ pulse sequence, displayed in Fig.~\ref{fig:infid_scaling} for several qudit dimensions. Note that some of the phases for non-prime $d$ in Eq.~\eqref{eq:CZ_general_qudit} are trivial (integer multiples of $2\pi$), leading to improved fidelities in those cases.

We may further obtain a simple approximate formula for the infidelity as a function of $d$. The inset of Fig.~\ref{fig:infid_scaling} shows the weighted number of non-trivial $\CRg$ pulses for each $d$, counting two-tone drives as two pulses due to the doubled population in the Rydberg levels. For prime $d$, the $\CZg$ pulse sequence involves $d-1$ single-tone pulses and $\frac{1}{2}(d-1)(d-2)$ two-tone pulses, amounting to a weighted total of $(d-1)^2$ pulses. The average duration of these $\CRg$ pulses, also shown in the inset of Fig.~\ref{fig:infid_scaling}, generally decreases very slowly with $d$. We thus predict the $\CZg$ infidelity to scale as
\begin{equation}\label{eq:infid_scaling}
    1 - F_{\CZg}\approx1 - \left(F_{\CRg_{\{1\}}(\pi)}\right)^{(d-1)^2},
\end{equation}
where $F_{\CRg_{\{1\}}(\pi)}$ is the fidelity of the gate that imparts a conditional phase $\pi$ to a single level (i.e., $\CZg$ for $d=2$). Let us stress that this formula does not take into account other error sources, such as shot-to-shot detuning, that could also affect the single-qudit operations required between the $\CRg$ pulses of Eq.~\eqref{eq:CZ_general_qudit}. However, these additional noise factors are typically weaker than Rydberg decay, which presents a fundamental lower bound on the gate infidelity.
% However, it is reasonable to assume that Rydberg decay is the dominant source of infidelity~\cite{peper_spectroscopy_2025}. Increasing the Rabi frequency should reduce the time spent in the Rydberg levels and therefore improve the performance of the gate with the infidelity scaling as a power law $1-F_{\CZg}\sim\OmMax^{-1}$~\cite{tsai_benchmarking_2025}.

% The fidelity of the $\CZg$ gate is roughly equal to the product of the $\CRg$ fidelities; indeed, we may crudely estimate how the $\CZg$ fidelity scales with the qudit dimension by counting the number of single- and two-tone pulses applied in Eq.~\eqref{eq:CZ_general_qudit}. In the qutrit case, we find $F_{\CZg} \approx F_{\CRg_{\{1\}}}F_{\CRg_{\{2\}}} F_{\CRg_{\{1,2\}}}\approx F_{\CRg_{\{2\}}}^4$, since $F_{\CRg_{\{1,2\}}}\approx F_{\CRg_{\{2\}}}^2$.

\subsection{Comparison with qubit-based implementations}
We conclude this section by comparing our framework with qubit-based implementation schemes of qudit gates. Indeed, an alternative strategy to ours is to bunch together multiple atoms, each realizing a single qubit, and use a subset of the available multi-qubit states to construct a single qudit. Our approach is clearly more scalable; for instance, we need half as many physical atoms to realize a system with a given number of qutrits, compared with qubit-based implementations, which require two physical atoms for a single logical qutrit. Scaling to larger $d$ would require an overwhelming overhead in the qubit case. Another significant advantage is achieved in single-qudit operations: qubit-based implementations require the use of two-qubit entangling gates for universal single-qudit control, leading to considerable degradation in fidelity. Our native approach avoids this issue.

With regards to qudit-entangling gates, we expect our framework to be at least on par with qubit-based schemes. As a concrete example, consider again the qutrit $\CZg$ gate, which we simulate with a fidelity of 0.994 (in the absence of Rydberg tone crosstalk). We compare this with the qubit-based protocol of Ref.~\cite{iqbal2024qutrittoriccodeparafermions} for realizing the ${\mathbb Z}_3$ toric code, where each qutrit is represented by two physical qubits. There, the qutrit $\CZg$ gate is realized using four operations that are each equivalent to our $\CRg_{\{1\}}\!\left(\frac{4\pi}{3}\right)$ gate. If we were to adapt the protocol of Ref.~\cite{iqbal2024qutrittoriccodeparafermions} to the neutral atoms platform with the same experimental conditions as those simulated in this work, we would achieve a fidelity of $0.998^4\approx0.994$, equal to our fidelity within statistical error (as mentioned above, that is to be expected, since the infidelity of the $\CRg_{\{1,2\}}\!\left(\frac{4\pi}{3}\right)$ gate is roughly twice that of the $\CRg_{\{1\}}\!\left(\frac{4\pi}{3}\right)$ gate due to the doubled average population in the Rydberg levels). We further note that the $\CZg$  protocol used in Ref.~\cite{iqbal2024qutrittoriccodeparafermions} requires several steps where specific pairs of qubits are locally isolated in order to implement the $\CRg_{\{1\}}\!\left(\frac{4\pi}{3}\right)$ gates (i.e., the required drives are not global), and would thus result in diminished fidelity. More importantly, let us stress that the preparation of the $\mathbb{Z}_3$ toric code requires additional steps other than the $\CZg$ gates. One needs to prepare qutrit states and apply single-qutrit gates, operations that involve entangling gates in qubit-based platforms. Our native qutrit scheme is considerably more suited to these operations. Therefore, the overall performance of the full $\mathbb{Z}_3$ toric code preparation protocol should be markedly better on our proposed device. % Moreover, our scheme would also provide superior performance for other entangling gates, such as $\mathrm{C}\mathcal{X}$, that can be reached from $\CZg$ using single-qudit rotations.

% As a concrete example, consider the protocol proposed in Ref.~\cite{iqbal2024qutrittoriccodeparafermions} which constructs qutrits out of pairs of qubits on a trapped ions platform, and realizes the $\CZg$ two-qutrit gate using 4 applications of the $\CRg\left(\frac{4\pi}{3}\right)$ gate. Our Eq.~\eqref{eq:CZ_general_qutrit} applies this gate 3 times, resulting in improved fidelity. Moreover, achieving universal single-qudit control in qubit-based implementations requires the application of qubit-entangling gates, whose fidelity is degraded by the Rydberg decay, as opposed to our native gates which do not involve the Rydberg levels.

% \section{Applications}
% Measurement-based prep of $S_3$

% magic state distillation?

% Quantum simulation 

% QEC?

\section{Summary and Outlook}\label{sec:summary}
The rich level structure of neutral atoms is naturally suited to the native encoding and control of qudits with local dimension $d>2$. Multiple stable and metastable states associated with each atom facilitate the coherent storage of information beyond the qubit paradigm. Furthermore, inter-level transitions can be individually addressed by leveraging the non-linearity of the energy spectrum or its selection rules, allowing for the manipulation of the stored information, as well as qudit entanglement and readout using ancillary atomic levels. Still, the higher local dimension of the encoded information means that a larger number of independent control knobs is necessary. The challenge, therefore, is to devise a strategy for operating these knobs in a manner that is efficient and robust to realistic experimental limitations.

In this work, we presented such a strategy, formulating a recipe for implementing a universal set of single-qudit and two-qudit gates, applicable to any $d$. Single-qudit operations are realized using multi-tone pulses with $d-1$ incommensurate frequencies. Two-qudit operations are applied through global addressing, and are constructed from multi-tone pulses of two types: pulses inducing transitions between qudit states (amounting to single-qudit rotations) and pulses inducing transitions to Rydberg states (generating entanglement). 

This multi-tone approach distinguishes our proposal from previous proposals for universal control of qudits in neutral atom arrays. The use of several simultaneous tones expedites the runtime of each gate, which is further optimized through quantum optimal control methods. The resulting pulses have simple amplitude and phase profiles, making them directly suitable for current hardware capabilities.

Given the importance of the two-qudit $\CZg$ gate, we provided a general protocol for its implementation. This protocol involves two-tone driving of Rydberg transitions, which, as we proved, is a minimal necessary component of a global $\CZg$ gate and, more generally, of global symmetric two-qudit phase gates, for $d>3$. Our work therefore constitutes a minimal scheme for implementing any entangling symmetric two-qudit phase gate of any qudit dimension, tailored to the strengths of existing hardware. We furthermore thoroughly analyzed the application of our strategy to the case of qutrits, deriving explicit pulse shapes that realize elementary gates ($\Xg$, Hadamard, and $\CZg$), and performing extensive noise simulations to examine the robustness of the two-qutrit $\CZg$ gate. We found that the $\CZg$ gate is realized with high fidelity and can be successfully modified to overcome potential crosstalk between Rydberg tones.

Our work thus offers a clear and viable blueprint for the experimental demonstration of multi-qudit operations in neutral atom arrays. This will pave the path toward the realization of qudit quantum error correction codes~\cite{Anwar_2014QuditToricCode} and magic state distillation protocols~\cite{Campbell2012, Campbell2014}. It will also open up enticing prospects for simulating unique quantum many-body effects, including salient phenomena in spin-1 systems such as symmetry-protected topological order~\cite{AKLT1987,Edmunds2025} and ergodicity-breaking dynamics~\cite{Schecter2019_Scars,Sala2020_HSF}, or pair creation and string breaking in the real-time dynamics of lattice gauge theories~\cite{Meth2025LGT,martinez_real-time_2016,klco_quantum_2018,LGT_review2020_truncated,zhou_thermalization_2022,Cochran2025NatureLGT_truncated,Gonzalez_Cuadra2025NatureLGT,halimeh_cold-atom_2025}.
The favorable flexibility and coherence time of neutral atom arrays would enable us to study such physics in various spatial dimensions and access long-time nonequilibrium dynamics. Our approach is particularly well-suited for the simulation of models where the physical time evolution involves state-dependent phase accumulation, which is the native two-qudit operation enabled by the platform. Such operations appear, e.g., in the context of simulating interacting spin systems and lattice gauge theories~\cite{Meth2025LGT, Gavreev2025}.

The control scheme we presented allows for versatility in the techniques used for driving elementary transitions, and it can be readily enhanced through methods that yield improved speed and noise resilience. For example, it was recently proposed that dipole-dipole interactions between neutral atoms can be leveraged to produce faster entangling gates compared with those based on the Rydberg blockade mechanism~\cite{Giudici2025FastEntangling}, a tactic to which our scheme can be seamlessly adapted. It is also possible to integrate other existing schemes of single-qudit control, such as the tensor light shift method of Refs.~\cite{Ahmed2025, Omanakuttan_2021_OptimalControl}, into our composite entangling gates. Another natural extension of our strategy is to implement entangling gates between more than two qudits, or between qudits of differing dimensions \cite{verresen2022efficiently,Gaz2025LGT_D8}, by imparting conditional phases on selected subspaces of Rydberg-coupled levels. More broadly, it will be interesting to consider whether our multi-tone approach can lead to similarly robust multi-qudit control schemes in other types of quantum platforms, such as those based on trapped ions or superconducting circuits.

\nocite{zenodo_repo}

\section*{Acknowledgments}
We thank 
Nathanan Tantivasadakarn and Erez Zohar for useful discussions.
A.B.~is supported by the Adams fellowship Program of the Israel Academy of Sciences and Humanities. S.F.~is supported by the Azrieli Foundation Fellows program. M.G.~is supported by the Israel Science  Foundation (ISF) and the Directorate for Defense Research and Development (DDR\&D) through Grant No.~3427/21, the ISF grant No.~1113/23, and the US-Israel Binational Science Foundation (BSF) through Grant No.~2020072. This research was supported by Israel Science Foundation research grant (ISF’s No.~4098/25) and the Maimonides Fund’s Future Scientists Center. We thank the Tel Aviv University Center of Light-Matter Interaction for their support.

\appendix
% \section{Universal gate sets for qudits with prime local dimension}\label{app:UniGateSet}

% We therefore assume throughout the rest of this section that $d$ is prime. 

% The gates $\Xg$ and $\Zg$ generate the Pauli group, defined as ${\cal P}=\left\{\omega^a\Xg^b\Zg^c| a,b,c=0,\ldots,d-1\right\}$. Tensor products of operators from ${\cal P}$, or ``Pauli strings", are used to define an important subgroup of unitary multi-qudit operators, namely the Clifford group. An $n$-qudit unitary operator $C$ is said to be in the Clifford group on $n$ qudits if, for any $U\in{\cal P}^{\otimes n}$, it satisfies $CUC^{\dagger}\in{\cal P}^{\otimes n}$. The magic-state model, a standard scheme for fault-tolerant quantum computation, relies on the system evolving solely through Clifford operations (in addition to the efficient preparation of appropriate initial states)~\cite{Bravyi2005,Campbell2012}, making the high-fidelity realization of Clifford operators an especially valuable goal.

% For any odd prime $d$, the entire $n$-qudit Clifford group can be generated using the single-qudit gates $\Hg$ and ${\Sg}=\prod_{k=0}^{d-1}\Rg_k\!\left(\frac{k(k+1)}{2}\right)$, together with the two-qudit entangling gate  $\CZg$~\cite{Clark_2006}, defined through

\section{Pulse optimization using quantum optimal control}\label{app:GRAPE}

In this appendix, we describe our method for obtaining pulse shapes that implement desired single-qudit or two-qudit operations. Given lasers that drive transitions either between qudit states or between a qudit state and a Rydberg level, our goal is to construct driving field functions that will realize a certain target gate in optimal time. We take the approach of quantum optimal control, largely following the path laid out by Ref.~\cite{Jandura2022} for neutral atom qubit systems.

Before providing full details, we briefly present the main principle of the method. Given a particular target gate $U_{\rm tar}$, we first choose the laser fields we would like to use for its realization, and set all other fields to zero (as a simple example: the fields that drive Rydberg transitions are not needed for single-qudit manipulations, and can be turned off). Let $\left\{\Omega_j(t)\right\}_{j=1}^M$ denote the non-zero laser fields, where the notation includes both types of fields (driving a transition either between qudit states or to a Rydberg level), and let $T$ denote the pulse duration. The functions $\left\{\Omega_j(t)\right\}_{j=1}^M$ will determine the unitary time-evolution operator $U_{\rm true}(T)$ under which the system will evolve, which should ideally be equal to $U_{\rm tar}$. The fidelity between $U_{\rm tar}$ and $U_{\rm true}(T)$ can be defined as~\cite{PEDERSEN2007}
\begin{equation}\label{eq:FidelityDef}
    F(T)=\frac{1}{d^{2n}}\left|{\rm Tr}\!\left[{U_{\rm tar}}^{\dagger}U_{\rm true}(T)\right]\right|^2,
\end{equation}
with $n$ being the number of qudits on which $U_{\rm tar}$ acts; we motivate this definition for $F(T)$ below.

We employ the Gradient Ascent Pulse Engineering (GRAPE) method~\cite{KHANEJA2004}, which relies on gradient descent in the space of possible complex functions $\left\{\Omega_j(t)\right\}_{j=1}^M$, to find the functions that maximize the fidelity $F(T)$. The gradient descent procedure is constrained by a bound on the Rabi frequencies, which cannot be larger than $\OmMax$. Once we have computed $F(T)$ for a range of $T$ values, we define the optimal pulse duration $T_{\rm opt}$ as that above which $F(T)\approx 1$, and below which $F(T)$ abruptly drops. The time-dependent fields obtained by the GRAPE procedure for $T=T_{\rm opt}$ yield the optimal-time pulse for the desired target gate.

In the remainder of this appendix, we explain in more detail the definition of the fidelity function $F(T)$ (Sect.~\ref{appsubsec:fidelity}); the GRAPE method for maximizing $F(T)$ (Sect.~\ref{appsubsec:GRAPE_discrete}); and an alternative version of the method, which employed Fourier sums to assure the smoothness of the resulting driving fields and reduce the computational runtime (Sect.~\ref{appsubsec:GRAPE_Fourier}).

\subsection{The fidelity function}\label{appsubsec:fidelity}

The definition in Eq.~\eqref{eq:FidelityDef} for the averaged gate fidelity originates in the following theorem, proved in Ref.~\cite{PEDERSEN2007}. Let $O$ be an operator acting on a $D$-dimensional Hilbert space, then its averaged squared expectation value is given by
\begin{equation}
    {\mathbb E}_{\ket{\psi}\sim{\rm Haar}}\!\left[\left|\bra{\psi}O\ket{\psi}\right|^2\right]=\frac{\left|{\rm Tr}\left(O\right)\right|^2+{\rm Tr}\left(OO^{\dagger}\right)}{D\left(D+1\right)},
\end{equation}
where the average is taken over random pure states with respect to the Haar measure~\cite{Mele2024introductiontohaar}. If we substitute $O={U_{\rm tar}}^{\dagger}U_{\rm true}(T)$ into this formula, it yields the average squared overlap between $U_{\rm tar}\ket{\psi}$ and $U_{\rm true}(T)\ket{\psi}$ for a random state $\ket{\psi}$ (an overlap which ideally is equal to 1), corresponding to a natural notion of gate fidelity.

In general, the evolution of our system of interest occurs in a Hilbert space that includes not only the computational qudit states but also Rydberg states. By definition, $U_{\rm tar}$ is a unitary operator on the computational subspace, while $U_{\rm true}(T)$ can become non-unitary under projection onto this subspace, if it happens to have non-diagonal matrix elements between that subspace and the Rydberg manifold. We should therefore see the notations for both operators as referring to their projection onto the computational subspace of $n$ qudits, and $\ket{\psi}$ as drawn from that subspace, with dimension $D=d^n$.

Let us initially assume that $U_{\rm true}(T)$ is unitary when projected onto the computational subspace. In that case, $O={U_{\rm tar}}^{\dagger}U_{\rm true}(T)$ is a unitary operator, and we necessarily have ${\rm Tr}\left(OO^{\dagger}\right)=d^n$, so the fidelity is effectively determined by the value of $\left|{\rm Tr}\left(O\right)\right|^2$, which can range between 0 and $d^{2n}$. This immediately gives rise to the definition of the fidelity given in Eq.~\eqref{eq:FidelityDef}.

This definition remains meaningful even if the projected $U_{\rm true}(T)$ is non-unitary. To see this, note that it must be unitary \emph{before} the projection, simply because it arises from the Hamiltonian evolution of the system. This entails that the projected operator satisfies
\begin{equation}
    {\rm Tr}\left({U_{\rm true}(T)}^{\dagger}U_{\rm true}(T)\right)\le d^n,
\end{equation}
with equality holding if and only if the projected operator is unitary (the expression on the left-hand side is equal to the sum of norms of matrix columns, which cannot increase under projection). Next, the Cauchy-Schwarz inequality (together with the unitarity of $U_{\rm tar}$) implies that
\begin{equation}
    \left|{\rm Tr}\left({U_{\rm tar}}^{\dagger}U_{\rm true}(T)\right)\right|^2\le d^n\left|{\rm Tr}\left({U_{\rm true}(T)}^{\dagger}U_{\rm true}(T)\right)\right|.
\end{equation}
This means that when $F(T)=1$ in Eq.~\eqref{eq:FidelityDef}, $U_{\rm true}(T)$ must be unitary, as desired for our optimization procedure.

\subsection{The GRAPE method}\label{appsubsec:GRAPE_discrete}
Given a fixed value of $T$, the GRAPE method~\cite{KHANEJA2004} allows us to maximize $F(T)$ within the space of allowed functions $\left\{\Omega_j(t)\right\}_{j=1}^M$. Importantly, it reduces the optimization problem to a gradient descent problem in a finite-dimensional parameter space, which can then be solved by standard numerical means.

First, let us introduce the notations $\ket{g_j},\ket{e_j}$ for each $j$, representing the two levels of the transition driven by $\Omega_j(t)$. We may write the time-dependent Hamiltonian using $2M$ time-dependent real functions $u_m(t)$ and $2M$ time-independent Hermitian operators $H_m$, namely
\begin{equation}\label{eq:ControlFunDef}
    H(t)=\sum_{m=1}^{2M}u_m(t)H_m,
\end{equation}
where $\Omega_j(t)=2u_{2j-1}(t)+2iu_{2j}(t)$, $H_{2j-1}=\ket{g_j}\bra{e_j}+{\rm h.c.}$, and $H_{2j}=i\ket{g_j}\bra{e_j}+{\rm h.c.}$ for $j=1,\ldots,M$. 

Next, the pulse duration $T$ is sliced into $N$ small time intervals of equal length, $\delta t=T/N$. Approximating $u_m(t)$ as piecewise-constant functions that are constant within each time interval, the time evolution operator can be written as
\begin{equation}\label{eq:Decomp_TrueU}
    U_{\rm true}(T) = U_NU_{N-1}\cdot\ldots\cdot U_1,
\end{equation}
where
\begin{equation}
    U_r=\exp\!\left[-i\delta t\sum_{m=1}^{2M}u_m(r\cdot \delta t)H_m\right].
\end{equation}
Let us define $u_{m,r}\equiv u_m(r\cdot\delta t)$, then $\{u_{m,r}\}$ are the $2N\!M$ real parameters that we vary in our search for a maximum point of the fidelity function $F(T)$. $N$ must be chosen to be large enough to allow smooth variations of the fields.

To reduce the maximization problem to a standard gradient descent problem, we need to compute the derivatives of $F(T)$ with respect to the optimization parameters. These are given, up to an immaterial multiplicative constant, by
\begin{equation}\label{eq:GRAPE_deriv}
    \frac{\partial F(T)}{\partial u_{m,r}} = \delta t\cdot{\rm Im}\!\left\{{\rm Tr}\!\left[{U_{\rm tar}}^{\dagger}U_{\rm true}(T)\right]^* {\rm Tr}\!\left[{U_{\rm tar}}^{\dagger}W_{m,r}\right]\right\},
\end{equation}
where we defined
\begin{equation}\label{eq:definition_W_mr}
    W_{m,r} = U_N\cdot\ldots\cdot U_{r+1}H_mU_r\cdot\ldots\cdot U_1.
\end{equation}
The $2N\!M$ derivatives in Eq.~\eqref{eq:GRAPE_deriv} define the gradient descent flow in the parameter space, and the values $\left\{u_{m,r}\right\}$ that maximize $F(T)$ can then be obtained by conventional numerical methods. These optimal values then yield the optimal control functions $u_m(t)$, which in turn yield the required driving fields $\Omega_j(t)$.

\subsection{Modified GRAPE method with Fourier decomposition}\label{appsubsec:GRAPE_Fourier}

A potential flaw of the GRAPE procedure is that it permits sharp discontinuous jumps in the optimal functions $\left\{\Omega_j(t)\right\}_{j=1}^M$ that it produces. Indeed, as the time discretization gets finer with larger $N$, unphysically rapid variations of the fields, with rates $N/T$, become allowed (this may also defy the assumption underlying the rotating wave approximation). 

It is therefore advantageous in some instances to modify the GRAPE algorithm so that its search for the optimal driving fields is restricted to functions that are slowly varying by their definition. This can be done by imposing on the control functions $u_m(t)$ in Eq.~\eqref{eq:ControlFunDef} the form of a finite sum of Fourier terms, featuring a range of allowed frequencies. That is, we replace the control functions with
\begin{equation}\label{eq:u_Fourier}
    u_m(t) =a_{m,0} +\sum_{k=1}^{\frac{1}{2}(K-1)}\left[a_{m,k}\cos(k\omega_0t)+b_{m,k}\sin(k\omega_0t)\right],
\end{equation}
where $\left\{a_{m,k},b_{m,k}\right\}$ are real parameters, $\omega_0$ is the smallest non-trivial frequency of the Fourier decomposition, and $K$ marks the number of independent terms in the sum, setting a cutoff frequency equal to $\frac{1}{2}(K-1)\omega_0$.

After determining the values of $\omega_0$ and $K$ according to physical considerations of the problem at hand, we may treat $\left\{a_{m,k},b_{m,k}\right\}$ as our optimization parameters and perform a similar procedure to that described in Sect.~\ref{appsubsec:GRAPE_discrete}. We again slice the pulse duration into $N$ equal-length time intervals, with $N$ large enough so that $K\omega_0\ll N/T$, meaning that the control functions $u_m(t)$ are approximately constant within each time interval. This ensures that we can write $U_{\rm true}(T)$ as a product of unitaries that are exponents of the instantaneous system Hamiltonian, as in Eq.~\eqref{eq:Decomp_TrueU}. The derivatives that define the gradient descent are then given by
\begin{widetext}
\begin{align}
    \frac{\partial F(T)}{\partial a_{m,k}} &= \delta t\cdot{\rm Im}\!\left\{{\rm Tr}\!\left[{U_{\rm tar}}^{\dagger}U_{\rm true}(T)\right]^* {\rm Tr}\!\left[{U_{\rm tar}}^{\dagger}\sum_{r=1}^N \cos(k\omega_0r\cdot\delta t) W_{m,r}\right]\right\},\nonumber\\
    \frac{\partial F(T)}{\partial b_{m,k}} &= \delta t\cdot{\rm Im}\!\left\{{\rm Tr}\!\left[{U_{\rm tar}}^{\dagger}U_{\rm true}(T)\right]^* {\rm Tr}\!\left[{U_{\rm tar}}^{\dagger}\sum_{r=1}^N \sin(k\omega_0r\cdot\delta t) W_{m,r}\right]\right\},
\end{align}
\end{widetext}
where $W_{m,r}$ was defined in Eq.~\eqref{eq:definition_W_mr}. The dimension of the parameter space is $KM$, i.e., it is typically much smaller than $2NM$, the dimension in the case of the algorithm described in Sect.~\ref{appsubsec:GRAPE_discrete}. Therefore, this modified version of the optimization method not only inherently imposes the smoothness of $\Omega_j(t)$, but also significantly reduces the dimensionality of the problem. This alternative method is indeed employed in Appendix~\ref{app:CZ_fourier}.

\section{Generating ${\rm SU}(d)$ using $d-1$ independent lasers}\label{app:SingleQuditControl}
Here we prove that the $d-1$ lasers described in Sect.~\ref{subsec:SingleQuditGate} can be used to apply any ${\rm SU}(d)$ operator to a single qudit in our system. Indeed, let us begin by considering only three levels $|0\rangle,|1\rangle,|2\rangle$. Suppose that we control two lasers that govern the qudit dynamics through the Hamiltonians ${\cal H}_{0,1}$ and ${\cal H}_{1,2}$, inducing the transitions $|0\rangle\leftrightarrow|1\rangle$ and $|1\rangle\leftrightarrow|2\rangle$, respectively [the definition of ${\cal H}_{j_1,j_2}$ appears in Eq.~\eqref{eq:RabiHamiltonian}]. Due to our full control of the Rabi and detuning frequencies of each lasers, the dynamics we may impose is equivalent to that dictated by the evolution operator ${\cal T}\exp\!\left[-i\int{\rm d}t\sum_{m=1}^4f_m(t)\lambda_m\right]$, where ${\cal T}$ is the time-ordering operator, $f_m(t)$ are arbitrary time-dependent real functions, and $\lambda_m$ are the following Hermitian and traceless matrices:
\begin{align}\label{eq:SU3Generator}
\lambda_1 & = \begin{pmatrix}0 & 1 & 0\\
1 & 0 & 0\\
0 & 0 & 0\end{pmatrix},\quad\lambda_2 = \begin{pmatrix}0 & -i & 0\\
i & 0 & 0\\
0 & 0 & 0\end{pmatrix},\nonumber\\
\lambda_3 & = \begin{pmatrix}0 & 0 & 0\\
0 & 0 & 1\\
0 & 1 & 0\end{pmatrix},\quad\lambda_4 = \begin{pmatrix}0 & 0 & 0\\
0 & 0 & -i\\
0 & i & 0\end{pmatrix}.
\end{align}
These are in fact the Pauli $\sigma_x$ and $\sigma_y$ operators in the relevant subspaces. By appropriately choosing $f_m(t)$, we may realize any unitary of the form $\exp\!\left[-i\sum_m\alpha_mC_m\right]$, where $\alpha_m$ are arbitrary real coefficients and $C_m$ are fixed Hermitian operators that include the four matrices in Eq.~\eqref{eq:SU3Generator} as well as operators proportional to the commutators of these matrices (thanks to the Baker-Campbell-Hausdorff formula).

We now observe that the commutators of these four matrices are proportional to four other traceless Hermitian matrices:
\begin{align}\label{eq:SU3GeneratorCommutator}
\frac{1}{2i}[\lambda_1,\lambda_2] & = \begin{pmatrix}1 & 0 & 0\\
0 & -1 & 0\\
0 & 0 & 0\end{pmatrix},\;\frac{1}{2i}[\lambda_3,\lambda_4] = \begin{pmatrix}0 & 0 & 0\\
0 & 1 & 0\\
0 & 0 & -1\end{pmatrix},\nonumber\\
\frac{1}{i}[\lambda_1,\lambda_3] & = \begin{pmatrix}0 & 0 & -i\\
0 & 0 & 0\\
i & 0 & 0\end{pmatrix},\;i[\lambda_1,\lambda_4] = \begin{pmatrix}0 & 0 & 1\\
0 & 0 & 0\\
1 & 0 & 0\end{pmatrix}.
\end{align}
The matrices in Eqs.~\eqref{eq:SU3Generator} and \eqref{eq:SU3GeneratorCommutator} constitute together a set of 8 linearly-independent, mutually-orthogonal (under the trace inner product), traceless Hermitian matrices. In general, a set of $d^2-1$ such matrices is sufficient in order to generate the group ${\rm SU}(d)$, and so these matrices generate ${\rm SU}(3)$.

The generalization of this statement to $d>3$ is then straightforward. A laser that induces the transition $|j_1\rangle\leftrightarrow|j_2\rangle$ provides us as generators the $\sigma_x$ and $\sigma_y$ operators in the subspace spanned by $\left\{|j_1\rangle,|j_2\rangle\right\}$, and when combined with a laser inducing the transition $|j_2\rangle\leftrightarrow|j_3\rangle$, we have as generators also the $\sigma_x$ and $\sigma_y$ operators in the subspace spanned by $\left\{|j_1\rangle,|j_3\rangle\right\}$. Therefore, if our $d-1$ lasers connect all $d$ levels of the qudit, then our generator set includes $\sigma_x$ and $\sigma_y$ within the subspace spanned by any choice of two levels; these are $d^2-d$ independent generators overall. To these we may add the $d-1$ operators $|0\rangle\langle0|-|j\rangle\langle j|$ with $1\le j \le d-1$ that the generator set includes due to the commutation relations of the various $\sigma_x$ and $\sigma_y$ operators. In total, this amounts to the set of $d^2-1$ independent generators that is necessary and sufficient for generating ${\rm SU}(d)$.

\section{Qutrit $\CZg$ gate using simultaneous single-qudit rotation and Rydberg drive} \label{app:CZ_fourier}
\begin{figure}[b]
    \centering
    \includegraphics[width=1\linewidth]{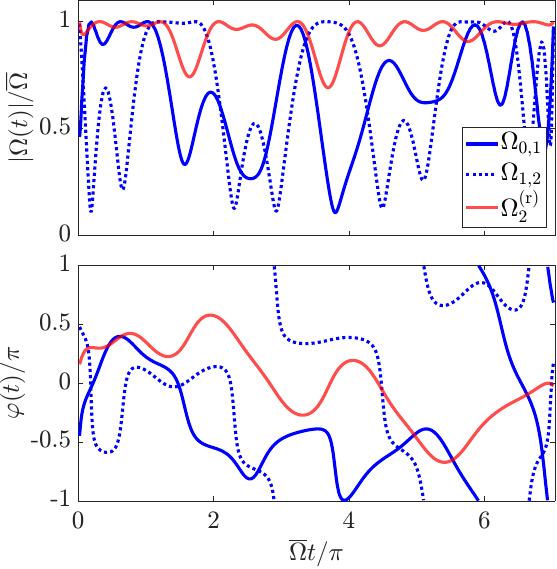}
    \caption{\textbf{Pulse shapes implementing the qutrit $\CZg$ gate,} obtained from GRAPE optimization in Fourier space. The apparent phase discontinuities are $2\pi$ phase wraps.}
    \label{fig:CZ_Fourier}
\end{figure}

An alternative to the pulse-based implementation in Eqs.~\eqref{eq:CZ_qutrit_single_ryd} and \eqref{eq:CZ_general_qutrit} is to simultaneously drive single-qudit and Rydberg tones. We use GRAPE in Fourier space (see Appendix~\ref{appsubsec:GRAPE_Fourier}) to find the laser amplitude and phase profiles that drive the $\ket{0}\leftrightarrow\ket{1}$, $\ket{1}\leftrightarrow\ket{2}$, and $\ket{2}\leftrightarrow\ket{r_2}$ transitions and implement the qutrit $\CZg$ gate. The result is displayed in Fig.~\ref{fig:CZ_Fourier}. Optimizing in Fourier space grants control over the smoothness of the resulting laser profiles, by setting the cutoff frequency $\frac{1}{2}(K-1)\omega_0$ defined in Eq.~\eqref{eq:u_Fourier}; in Fig.~\ref{fig:CZ_Fourier}, we choose $\frac{1}{2}(K-1)\omega_0 = 3\OmMax$ and $\omega_0=\OmMax/2$ (such that $K=13$), where $\OmMax$ is the maximal Rabi frequency (assumed to be the same for all three lasers). While the laser profiles in Fig.~\ref{fig:CZ_Fourier} may be smooth enough for practical implementation, realistic systematic error sources could considerably degrade the fidelity of the gate, and calibration would be much more challenging than the calibration needed for the pulsed versions in Eqs.~\eqref{eq:CZ_qutrit_single_ryd} and \eqref{eq:CZ_general_qutrit}.

\section{Proof of no-go theorem on implementing $\CZg$ using a single Rydberg transition}\label{app:CZ_no_go}
Here we characterize the class of two-qudit phase gates that can be implemented as composite pulse sequences involving global drives and only one Rydberg transition being resonantly driven. Specifically, we prove that, for any two-qudit unitary operator of the form given in Eq.~\eqref{eq:OneRydPhaseGate}, its action on the computational basis states must follow Eq.~\eqref{eq:PhaseOpNoGo}.

Writing $\CRg_{\{2\}}(\theta_k)={{\mathbb I}_d}^{\otimes 2}+(e^{i\theta_k}-1)\ket{2,2}\bra{2,2}$ and substituting this into Eq.~\eqref{eq:OneRydPhaseGate}, we obtain
\begin{widetext}
\begin{align}
    U=&\,W^{\otimes 2}+\sum_{k=1}^n(e^{i\theta_k}-1)\left(W_n\ldots W_k\right)^{\otimes 2}\ket{2,2}\bra{2,2}\left(W_{k-1}\ldots W_0\right)^{\otimes 2}\nonumber\\
    &+\sum_{1\le k_1 < k_2\le n}(e^{i\theta_{k_2}}-1)(e^{i\theta_{k_1}}-1)\left(W_n\ldots W_{k_2}\right)^{\otimes 2}\ket{2,2}\bra{2,2}\left(W_{k_2-1}\ldots W_{k_1}\right)^{\otimes 2}\ket{2,2}\bra{2,2}\left(W_{k_1-1}\ldots W_0\right)^{\otimes 2}\nonumber\\
    &+\ldots,
\end{align}
where we defined $W=W_nW_{n-1}\ldots W_0$, and where the ellipsis stands for sums of similar expressions, only with more than two insertions of $\ket{2,2}\bra{2,2}$ into the product of $\left(W_k\right)^{\otimes 2}$. Terms of the form $\bra{2,2}\left(W_{k_2-1}\ldots W_{k_1}\right)^{\otimes 2}\ket{2,2}$ are simply fixed matrix elements, so we observe that we may equivalently write
\begin{equation}
    U=W^{\otimes 2}+\sum_{0\le k_1 < k_2\le n}C_{k_1,k_2}\left(W_n\ldots W_{k_2}\right)^{\otimes 2}\ket{2,2}\bra{2,2}\left(W_{k_1}\ldots W_0\right)^{\otimes 2},
\end{equation}
\end{widetext}
where $C_{k_1,k_2}$ are some complex numbers. We further notice that terms of the form $\bra{j,m}\left(W_n\ldots W_{k_2}\right)^{\otimes 2}\ket{2,2}$ or $\bra{2,2}\left(W_{k_1}\ldots W_0\right)^{\otimes 2}\ket{j,m}$ are invariant under the exchange of $j$ and $m$. This means that, for any two-qudit state $\ket{\psi}$, we obtain that
\begin{align}
    U\ket{\psi}=&\,W^{\otimes 2}\ket{\psi}+\sum_{j=0}^{d-1}f_{\{j\}}(\ket{\psi})\ket{j,j}\nonumber\\
    &+\sum_{0\le j < m \le d-1}f_{\{j,m\}}(\ket{\psi})\left[\ket{j,m}+\ket{m,j}\right],
\end{align}
where $f_{\{j\}}$ and $f_{\{j,m\}}$ are linear functionals (i.e., linear maps from ${\mathbb C}^{d^2}$ to ${\mathbb C}$) that are symmetric under qudit exchange. In particular, these functionals vanish when they are applied to any two-qudit state that is anti-symmetric under qudit exchange.

Now, recall that $U$ is assumed to be diagonal in the computational basis, and let us define the phase $\xi_{0,1}$ via $U\ket{0,1}=e^{i\xi_{0,1}}\ket{0,1}$. Projecting both sides of this equality onto either $\ket{0,1}$ or $\ket{1,0}$ and subtracting the two equations yields
\begin{equation}\label{eq:phase_01}
    e^{i\xi_{0,1}} = \bra{0}W\ket{0}\bra{1}W\ket{1} - \bra{1}W\ket{0}\bra{0}W\ket{1}.
\end{equation}
Additionally, we use the fact that the linear functionals $f_{\{j\}}$ and $f_{\{j,m\}}$ vanish when applied to $\ket{0,1}-\ket{1,0}$ to see that $U\left(\ket{0,1}-\ket{1,0}\right)=W^{\otimes 2}\left(\ket{0,1}-\ket{1,0}\right)$. Projecting both sides onto either $\ket{0,2}$ or $\ket{1,2}$, we observe that
\begin{align}
    0& =\bra{0}W\ket{0}\bra{2}W\ket{1}-\bra{0}W\ket{1}\bra{2}W\ket{0},\nonumber\\
    0& =\bra{1}W\ket{0}\bra{2}W\ket{1}-\bra{1}W\ket{1}\bra{2}W\ket{0}.
\end{align}
Multiplying the first equality by $\bra{1}W\ket{1}$ and the second by $\bra{0}W\ket{1}$ and subtracting, we obtain
\begin{equation}
    0=e^{i\xi_{0,1}}\bra{2}W\ket{1},
\end{equation}
where we substituted Eq.~\eqref{eq:phase_01}. This, of course, entails that $\bra{2}W\ket{1}=0$.

The same exact argument can be applied to any other off-diagonal element of the unitary $W$, meaning that $W$ is necessarily diagonal with respect to the single-qudit computational basis. Letting $e^{i\xi_j}=\bra{j}W\ket{j}$ denote its diagonal elements, we obtain Eq.~\eqref{eq:PhaseOpNoGo} as a straightforward generalization of Eq.~\eqref{eq:phase_01}. While Eq.~\eqref{eq:PhaseOpNoGo} limits the eigenvalues of the states $\ket{j,m}$ with $j\neq m$, the eigenvalues of the states $\ket{j,j}$ can be made arbitrary; indeed, $\CRg_{\{2\}}(\theta)$ can be used to assign an arbitrary phase to the state $\ket{2,2}$, and, through single-qudit swaps of the states $\ket{2}$ and $\ket{j}$, also to any state $\ket{j,j}$ with $j\neq2$.

\section{Simulation in the presence of Rydberg tone crosstalk}\label{app:crosstalk}
\begin{figure}
    \centering
    \includegraphics[width=1\columnwidth]{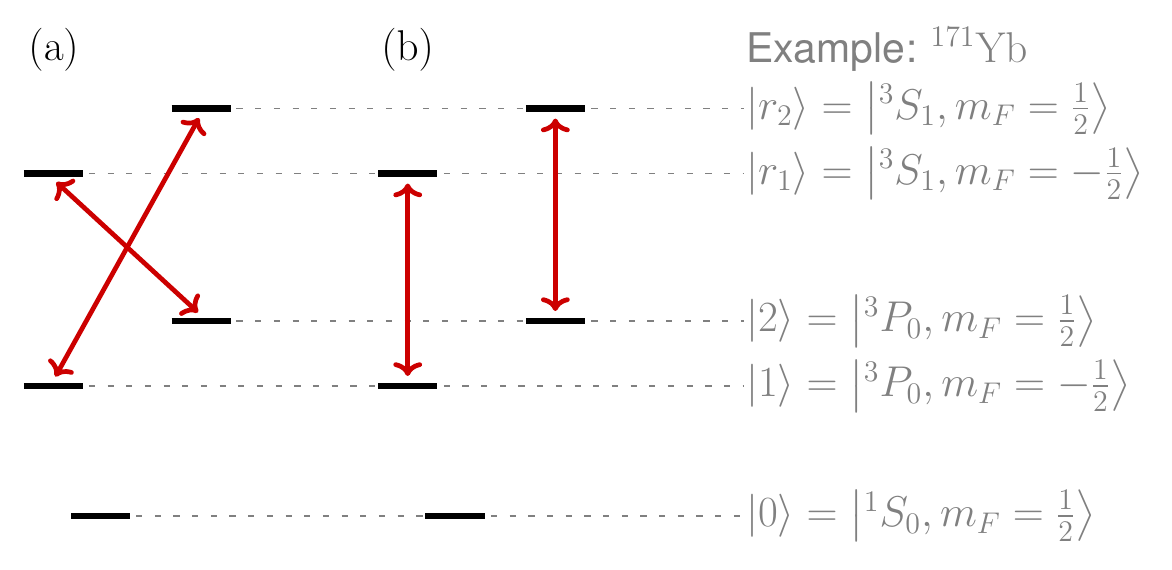}
    \caption{\textbf{Possible polarizations of the Rydberg tones.} \textbf{(a)} Circular polarization in the $F=1/2$ manifold, eliminating Rydberg tone crosstalk. \textbf{(b)} Linear polarization, inducing crosstalk between the $\ket{1}\leftrightarrow\ket{r_1}$ and $\ket{2}\leftrightarrow\ket{r_2}$ transitions. The difference between these transition frequencies is $\delta\omega_{1,2}$ in Eq.~\eqref{eq:H_drive_lightshift}.}
    \label{fig:polarizations}
\end{figure}

The use of circularly-polarized laser beams, depicted in Fig.~\ref{fig:polarizations}(a), eliminates crosstalk between the clock-to-Rydberg tones. However, the effect of such crosstalk needs to be addressed if the use of a single frequency-modulated beam is preferred. In this appendix, we show how to alleviate the error induced by the crosstalk by modifying the pulse shapes used within the scheme proposed in Eq.~\eqref{eq:CZ_general_qudit}. We then repeat the simulation of the $\CZg$ gate. To compare with the results of Sect.~\ref{sec:Imperfections}, we focus again on $d=3$, although our strategy to modify the pulse shapes applies at any qudit dimension.

As mentioned above, our protocol for the $\CZg$ gate assumes that the transitions to the Rydberg levels may be driven independently. This could be incompatible with the use of linearly-polarized lasers, illustrated in Fig.~\ref{fig:polarizations}(b); driving the transition $\ket{1}\leftrightarrow\ket{r_{1}}$ could also drive the transition $\ket{2}\leftrightarrow\ket{r_{2}}$, if the Rabi frequency of the drive is comparable with the detuning of the two transitions. The drive Hamiltonian reads

\begin{align}\label{eq:H_drive_lightshift}
    H_{1,r_{1}}  &= \frac{\Omega_{1}^{(\mathrm{r})}(t)}{2}\left(\ket{1}\bra{r_{1}}\!\otimes\!{\mathbb I}+{\mathbb I}\!\otimes\!\ket{1}\bra{r_{1}}\right) \nonumber \\
    &+\frac{\Omega_{1}^{(\mathrm{r})}(t)e^{i\delta\omega_{1,2}t}}{2}(\ket{2}\bra{r_{2}}\!\otimes\!{\mathbb I}
    +{\mathbb I}\!\otimes\!\ket{2}\bra{r_{2}})+{\rm h.c.}\quad\quad\quad
\end{align}
Here $\mathbb{I}$ is the identity matrix of a single atom, and $\delta\omega_{1,2}$ is the difference between the $1$ and $2$ transition frequencies, resulting from the finite Zeeman splitting in the Rydberg manifold. Note that the transition $\ket{1}\leftrightarrow\ket{r_2}$ is forbidden by selection rules (the two states have differing values of $m_F$, as illustrated in Fig.~\ref{fig:polarizations}), and is therefore absent from Eq.~\eqref{eq:H_drive_lightshift}.

% relying on the assumption that the Rabi frequencies are much smaller than the difference between the transition frequencies. The level spacing in the Rydberg manifold is due to a Zeeman splitting induced by an external magnetic field; in practice, this splitting is finite, and could be not very large compared with the Rabi frequencies. This leads to light shifts which could impact the fidelity of the gate. The Hamiltonian that accounts for these unwanted transitions reads

\begin{comment}
\begin{widetext}
\begin{align}\label{eq:H_drive_lightshift}
    H
    =&\,\,\sum_{j=1}^{d-1}\left[\frac{\Omega_j^{({\mathrm r})}(t)}{2}\left(|j\rangle\langle r_j|\!\otimes\!{\mathbb I}_{2d-1}+{\mathbb I}_{2d-1}\!\otimes\!|j\rangle\langle r_j|\right)+{\rm h.c.}\right]+ \sum_{j_1,j_2=1}^{d-1}V_{j_1,j_2}|r_{j_1}\rangle|r_{j_2}\rangle\langle r_{j_1}|\langle r_{j_2}|\nonumber\\
    &+ \sum_{\underset{j_1\neq j_2}{j_1,j_2=1}}^{d-1}\left[\frac{\Omega_{j_1}^{({\mathrm r})}(t)e^{i\delta \omega_{j_1,j_2}}}{2}\left(|j_2\rangle\langle r_{j_2}|\!\otimes\!{\mathbb I}_{2d-1}+{\mathbb I}_{2d-1}\!\otimes\!|j_2\rangle\langle r_{j_2}|\right)+{\rm h.c.}\right],
\end{align}
\end{widetext}
\end{comment}

If $\delta\omega_{1,2} \gg \Omega_{1}^{(\mathrm{r})}$, the unwanted transition term could be omitted within the rotating wave approximation. To lowest order in perturbation theory, driving the transition $\ket{1}\leftrightarrow\ket{r_{1}}$ shifts the energies of $\ket{2}$ and $\ket{r_{2}}$ by $\left|\Omega^{(\mathrm{r})}_{1}\right|^2/\delta \omega_{1,2}$. However, at small enough $\delta\omega_{1,2}$, a perturbative analysis is invalid and the unwanted coupling term could drastically alter the dynamics. Assuming the realistic values $\left|\Omega^{(\mathrm{r})}_{1}\right|\sim 2\pi\times5~\textrm{MHz}$ and $\delta\omega_{1,2}\sim 2\pi\times50~\textrm{MHz}$~\cite{peper_spectroscopy_2025} and using the optimal pulse shape shown in Fig.~\ref{fig:cz3_pulsed}(b), we find that the fidelity of the $\CRg$ gates is reduced by as much as 0.1.

\begin{figure}[t]
    \centering
    \includegraphics[width=1\linewidth]{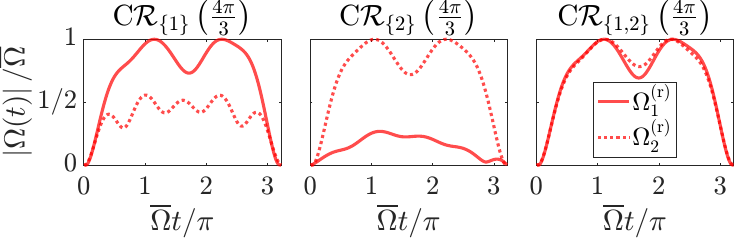}
    \caption{\textbf{Modified $\CRg\left(\frac{4\pi}{3}\right)$ gates,} with pulse shapes accounting for a finite Zeeman splitting of $\delta\omega_{1,2} = 2\pi\times50~\textrm{MHz}$, using a maximal Rabi frequency of $\OmMax = 2\pi\times5~\textrm{MHz}$.}
    \label{fig:CR_lightshift}
\end{figure}

To accommodate the finite level splitting, we optimize the pulse shapes for the $\CRg$ gates using the dynamics of the Hamiltonian in Eq.~\eqref{eq:H_drive_lightshift} in the GRAPE algorithm. We also account for the finite Rydberg interaction strength in the optimization. The pulse shapes are optimized in Fourier space to avoid sharp phase jumps that could not be implemented in practice (see Appendix~\ref{appsubsec:GRAPE_Fourier} for details). The amplitudes of the resulting pulse shapes implementing $\CRg_{\{1\}}\!\left(\frac{4\pi}{3}\right)$, $\CRg_{\{2\}}\!\left(\frac{4\pi}{3}\right)$, and $\CRg_{\{1,2\}}\!\left(\frac{4\pi}{3}\right)$, for a maximal Rabi frequency $\OmMax = 2\pi\times 5~\textrm{MHz}$ and Zeeman splitting $\delta\omega_{1,2} = 2\pi\times 50~\textrm{MHz}$, are displayed in Fig.~\ref{fig:CR_lightshift}. These pulses are slightly longer than the ideal pulse shape in Fig.~\ref{fig:cz3_pulsed}(b). In the implementation of $\CRg_{\{j\}}\!\left(\frac{4\pi}{3}\right)$, the dominant laser operates at near-maximal intensity, while the other laser operates at lower intensity to counteract the induced unwanted transition. Furthermore, the phase of the dominant laser (not shown in Fig.~\ref{fig:CR_lightshift}) roughly follows the profile of the optimal-time pulse in Fig.~\ref{fig:cz3_pulsed}(b).

Using the modified pulses in the presence of the finite Zeeman splitting, we find a fidelity of 0.993 for the $\CZg$ gate, slightly worse than the fidelity in the absence of crosstalk, 0.994. The degraded fidelity could be attributed to the increased population in the Rydberg levels induced by the crosstalk and the counteracting transitions driven in the implementation of $\CRg_{\{1\}}\!\left(\frac{4\pi}{3}\right)$ and $\CRg_{\{2\}}\!\left(\frac{4\pi}{3}\right)$.

\section{Effective population of the Rydberg manifold during a unitary evolution}\label{app:Rydberg_Time}
Any two-qudit entangling unitary operator is implemented through a unitary evolution $U(t)$ of two atoms for a given time $T$. This evolution necessarily involves transitions between the computational qudit states and ancillary Rydberg states, where the latter are prone to decay processes. To approximate the average Rydberg decay during such an evolution, we may estimate the effective portion of time out of the \textit{perfect} unitary evolution $U(t)$ during which the Rydberg states are populated, and then multiply this effective time by a characteristic decay rate of a Rydberg state.

Here we present a useful formula estimating the average population $\chi_{\rm ryd}$ of Rydberg states during the evolution $U(t)$ (the effective time will then be $\chi_{\rm ryd}T$). Let us first define the projector onto the computational two-qudit subspace,
\begin{equation}
    \Pi_{\rm comp}=\sum_{j,m=0}^{d-1}\ket{j,m}\bra{j,m}.
\end{equation}
We also define the projectors
\begin{align}
    \Pi_{\rm ryd} & =\sum_{j=0}^{d-1}\sum_{k=1}^{n_{\rm ryd}}\left[\ket{j,r_k}\bra{j,r_k}+\ket{r_k,j}\bra{r_k,j}\right],\nonumber\\
    \Pi_{\rm bloc}&=\sum_{k,l=1}^{n_{\rm ryd}}\ket{r_k,r_l}\bra{r_k,r_l},
\end{align}
where $n_{\rm ryd}$ is the number of Rydberg states that are connected to the qudit states by the underlying Hamiltonian. Respectively, these projectors perform projections onto the subspace spanned by states where a Rydberg state is occupied in only one of the atoms and the (blockaded) subspace of states with Rydberg states occupied by both atoms.

Suppose now that $\ket{\psi}$ is an arbitrary two-qudit state, i.e., $\Pi_{\rm comp}\ket{\psi}=\ket{\psi}$. A natural definition for the average population of the Rydberg manifold during the evolution $U(t)$ starting from the initial state $\ket{\psi}$ is
\begin{equation}\label{eq:RydTime}
    \chi_{\rm ryd}= \frac{1}{T}\int_0^T{\rm d}t\,\bra{\psi}U^{\dagger}\!\left(t\right)\left(\Pi_{\rm ryd}+2\Pi_{\rm bloc}\right)U\!\left(t\right)\ket{\psi},
\end{equation}
where the factor of 2 before $\Pi_{\rm bloc}$ stems from the fact that, in the corresponding subspace, both atoms are in a Rydberg state and the contribution to the Rydberg decay should be doubled.

This estimate currently depends on the state $\ket{\psi}$. We may eliminate this dependence by regarding it as a random two-qudit state and averaging over such states. We write $\ket{\psi}=\Pi_{\rm comp}V\ket{0,0}$ where $V$ is a Haar-random unitary operator within the two-qudit subspace, and use the rules of the Haar measure~\cite{Mele2024introductiontohaar} to find that
\begin{align}
    {\mathbb E}_{V\sim{\rm Haar}}\!\left[V^{\dagger}\Pi_{\rm comp}U^{\dagger}\!\left(t\right)\left(\Pi_{\rm ryd}+2\Pi_{\rm bloc}\right)U\!\left(t\right)\Pi_{\rm comp}V\right]\nonumber\\
    =\frac{1}{d^2}{\rm Tr}\!\left[U^{\dagger}\!\left(t\right)\left(\Pi_{\rm ryd}+2\Pi_{\rm bloc}\right)U\!\left(t\right)\Pi_{\rm comp}\right]{\mathbb I}.
\end{align}
Substituting this into the integrand in Eq.~\eqref{eq:RydTime}, we obtain the following formula for the average Rydberg population:
\begin{equation}\label{eq:RydTimeAvg}
    \chi_{\rm ryd}= \int_0^T\frac{{\rm d}t}{d^2T}\,{\rm Tr}\!\left[U^{\dagger}\!\left(t\right)\left(\Pi_{\rm ryd}+2\Pi_{\rm bloc}\right)U\!\left(t\right)\Pi_{\rm comp}\right].
\end{equation}

% The expression can be further simplified in the case of a perfect (infinite) Rydberg blockade. Then, the subspace represented by the projector $\Pi_{\rm bloc}$ cannot be accessed during the evolution, and we may drop its contribution to Eq.~\eqref{eq:RydTimeAvg} and replace $\Pi_{\rm ryd}$ with $1-\Pi_{\rm comp}$, leading to
% \begin{equation}
%     \chi_{\rm ryd}= 1-\int_0^T\frac{{\rm d}t}{d^2T}\,{\rm Tr}\!\left[U^{\dagger}\!\left(t\right)\Pi_{\rm comp}U\!\left(t\right)\Pi_{\rm comp}\right].
% \end{equation}

\bibliography{Qutrit_Optimal_Control}

\end{document}